\documentclass[journal]{IEEEtran}
\usepackage{amsmath,amsfonts}
\usepackage{array}
\usepackage[ruled]{algorithm2e}
\usepackage[noend]{algpseudocode}
\usepackage{textcomp}
\usepackage{stfloats}
\usepackage{url}
\usepackage{verbatim}
\usepackage{graphicx}
\usepackage{cite}
\usepackage{multirow}
\usepackage{makecell}
\usepackage{caption}
\usepackage{float} 
\usepackage{verbatim}
\usepackage{subfig}

\begin{document}
\title{Cross-Layer Optimization for Fault-Tolerant Deep Learning}

\author{
Qing Zhang, 
Cheng Liu,~\IEEEmembership{Member,~IEEE}, 
Bo Liu,
Haitong Huang, 
Ying Wang,~\IEEEmembership{Member,~IEEE}, \\
Huawei Li,~\IEEEmembership{Senior Member,~IEEE}, 
Xiaowei Li,~\IEEEmembership{Senior Member,~IEEE}
\thanks{This work was supported in part by the National Natural Science Foundation of China under Grant No. 62174162 and No. 62090024, and the Space Trusted Computing and Electronic Information Technology Laboratory of Beijing Institute of Control Engineering (BICE) under Grant OBCandETL-2022-07. \textit{(Corresponding author: Cheng Liu.)}}
\thanks{Qing Zhang, Cheng Liu, Haitong Huang, Ying Wang, Huawei Li, and Xiaowei Li are with both State Key Lab of Processors, Institute of Computing Technology, Chinese Academy of Sciences, Beijing 100190, China, and University of Chinese Academy of Sciences, Beijing 100190, China. (e-mail: \{zhangqing22s, liucheng\}@ict.ac.cn)}
\thanks{Bo Liu is with Beijing Institite of Control Engineering, Beijing 100190, China.}
}

\markboth{Journal of \LaTeX\ Class Files,~Vol.~14, No.~8, August~2024}%
{Shell \MakeLowercase{\textit{et al.}}: A Sample Article Using IEEEtran.cls for IEEE Journals}


\maketitle

\begin{abstract}
Fault-tolerant deep learning accelerator is the basis for highly reliable deep learning processing and critical to deploy deep learning in safety-critical applications such as avionics and robotics. Since deep learning is known to be computing- and memory-intensive, traditional fault-tolerant approaches based on redundant computing will incur substantial overhead including power consumption and chip area. To this end, we propose to characterize deep learning vulnerability difference across both neurons and bits of each neuron, and leverage the vulnerability difference to enable selective protection of the deep learning processing components from the perspective of architecture layer and circuit layer respectively. At the same time, we observe the correlation between model quantization and bit protection overhead of the underlying processing elements of deep learning accelerators, and propose to reduce the bit protection overhead by adding additional quantization constrain without compromising the model accuracy. Finally, we employ Bayesian optimization strategy to co-optimize the correlated cross-layer design parameters at algorithm layer, architecture layer, and circuit layer to minimize the hardware resource consumption while fulfilling multiple user constraints including reliability, accuracy, and performance of the deep learning processing at the same time. 
\end{abstract}


\section{Introduction}
\IEEEPARstart{D}{EEP} learning is continuously penetrating into broader application fields beyond traditional areas such as computer vision and natural language processing. More and more applications involve safety-critical scenarios\cite{ref1,ref2,ref5,ref8,ref9,ref22,ref23,ref29}, such as autonomous driving, aerospace, and robotics.Compared with conventional deep learning applications, safety-critical deep learning applications have additional requirements for reliability on top of accuracy and speed. Moreover, reliability has become a key indicator for such deep learning applications, directly determining whether they can be accepted for use. For example, the on-board deep learning module for vehicles must meet the reliability standards specified in ISO26262 in order to be applied to the vehicle system\cite{ref1,ref5,ref8}. As the core dedicated processor supporting high-energy-efficient deep learning processing\cite{ref11,ref13}, the deep learning accelerator (DLA), accompanied by higher transistor integration and lower threshold voltage in advanced nano-level processes, will inevitably be affected by soft faults, leading to abnormal errors in deep learning inference and even causing safety accidents\cite{ref3,ref7}.Therefore, designing fault-tolerant DLAs to effectively tolerate faults and achieve high-reliable inference is crucial to driving the application of deep learning in safety-critical fields.

The processing of deep learning is a typical compute- and memory-intensive task. Although traditional chip reliability design methods such as triple modular redundancy (TMR) can solve the problem of accuracy loss caused by faults,but, a large amount of redundant computation can greatly affect the processing speed, chip area, and power consumption of deep learning. This conflicts with the fundamental design goals of deep learning accelerators, such as speed and energy efficiency, also limits their applications in safety-critical areas.On the other hand, the outputs of deep learning processing are typically discrete, and many small computational errors have little impact on the inference results of deep learning. Additionally, deep learning processing involves many non-linear functions, which can filter out many intermediate calculation errors and mitigate the impact of soft errors on inference results. These features make deep learning processing more fault-tolerant compared to general-purpose computing. Many reliability design works for deep learning accelerators (DLA) utilize this natural fault-tolerant property of deep learning to reduce the fault-tolerant cost of deep learning, achieving a joint optimization of accuracy, reliability, and performance.

One major approach is to use selective redundancy protection based on the differential sensitivity of different parts of deep learning models to faults, thereby reducing the cost of fault tolerance without compromising the reliability and accuracy of deep learning computations. C. Schorn et al.\cite{ref26}  and Liu et al. \cite{ref3,ref12} propose to divide the computation array of DLA into different reliability regions for processing neurons of different importance, so that corresponding fault-tolerant protection can be applied to different regions to reduce the cost of fault tolerance. H. R. Mahdiani et al. \cite{ref31} propose to focus on protecting the high-order calculation part of multipliers to enhance the overall reliability of DLA. Some other works directly apply selective redundant computation at the model or algorithm level \cite{ref6,ref14,ref28,ref38,ref39} to improve the fault tolerance of deep learning and reduce the cost of fault tolerance.These works indicate that we can explore the inherent fault-tolerance differences in deep learning algorithms, accelerator architectures, circuits provide more protection for vulnerable parts to reduce fault-tolerance costs. Since different abstraction levels involve different information, there are significant differences in the corresponding fault-tolerance effects and costs. The circuit layer can accurately improve the fault tolerance of overall design but lacks flexibility and is difficult to directly use information from the application and architecture layers, and the fault tolerance cost is often relatively high. The algorithm layer can make better use of the information in the model layer, customize fault-tolerance choices for specific models, and reduce fault-tolerance costs but lacks low-level details. In cases where the Fault Rate is high and reliability requirements are also high, the fault-tolerance cost will increase sharply. The fault tolerance characteristics of the architecture layer are between the circuit and algorithm layers, with only some algorithm and circuit layer information ,but  still lacking a unified fault-tolerant deep learning architecture design. Previous DLA fault tolerance usually improved the reliability of DLA from one or two layers, lacking a system of cross-layer collaborative optimization and unable to fully utilize the advantages of different layer fault-tolerance designs. In addition, the relationships between cross-layer parameters are complex, the design space is large and manual design is not only inefficient and error-prone but also difficult to achieve global optimization, and it is challenging to quickly provide optimization solutions for different application requirements.

To address the above issues, we propose a cross-layer optimization design framework for fault-tolerant DLAs. At the model level, we analyze the sensitivity of deep learning computations to soft faults from two dimensions, neuron dimension and bit dimension. We categorize deep learning computation tasks into important neuron computations and ordinary neuron computations, and further divide neuron data into important bit and ordinary bit data to guide fine-grained selective protection of fault-tolerant DLA. At the architecture level, we propose a heterogeneous computing framework based on recomputing to process important neuron computations and common neuron computations separately, which supports selective protection of neuron computations. At the circuit level, we propose a bit-wise protected circuit redundancy design that only protects the important bit logical of the computing unit, and provides different circuit level redundancy protection designs for two types of computing arrays based on the sensitivity analysis of the algorithm layer, further reducing redundancy costs. In addition, we propose for the first time to reduce the cost of circuit-level bit-wise redundancy protection by quantization constraint, which limits the direct association logic size of important bit of deep learning neurons. Finally, according to the requirements of the target application, we automate the design selection of different layers and optimize the reliability, resource consumption, and performance of fault-tolerant DLA design.

The contributions of this work are summarized as follows:
\begin{enumerate}
    \item We analyze the sensitivity of deep learning models to soft faults in two dimensions: different neurons and Different bits of neurons. Based on this analysis, we propose fault-tolerant design of DLA across circuit, architecture, and algorithm levels, while optimizing multiple design goals including reliability, resource consumption, and performance.
    \item We propose for the first time the combination of a quantization strategy using deep learning models and bit redundancy design in DLA computing units. By constraining the quantization choices of the model, we can reduce the size of high-bit circuits that need protection without sacrificing accuracy. This leads to a lower cost of redundancy design at the circuit level for DLA.
    \item We have designed a fault-tolerant heterogeneous deep learning accelerator architecture based on recomputation. This architecture achieves fine-grained selective protection at the neuron level by taking into account the sensitivity differences of neurons to soft errors. It also tolerates the distribution differences of important neurons, reducing fault tolerance costs while ensuring computational accuracy.
    \item The experiments demonstrate that the cross-layer optimization design method proposed in this paper exhibits significant advantages in terms of reliability, resource consumption, and overall performance in DLA, compared to selective fault-tolerant designs at the individual circuit layer, architecture layer, and algorithm layer.
\end{enumerate}

\section{Related work}
Fault-tolerant design of DLA is the foundation to ensure the reliability of deep learning inference. In order to improve the reliability of DLA, many previous works have proposed various fault-tolerant design methods at different levels of abstraction such as circuit, architecture, and algorithm. We will introduce the relevant research works separately.

At the circuit level, traditional fault-tolerant coding techniques such as Error Correction Code (ECC) can be used to protect on-chip caches, and redundancy can protect control and computational logic. In order to reduce the cost of direct redundancy, H.R. Mahdiani et al. \cite{ref31} proposed giving higher priority protection to the high-order bits of deep learning computational units while ignoring the logic of low-bit positions to reduce the cost of redundant protection. Some works\cite{ref40,ref41,ref42}  proposed using logic circuits such as random computing and approximate computing to replace traditional binary computing logic to enhance fault tolerance or improve computational energy efficiency. Brandon Reagen et al. \cite{ref10} proposed using Razor circuits to detect timing faults caused by voltage drops, and then using the inherent fault-tolerance properties of deep learning models to add word and bit masks to repair SRAM data errors, thereby improving computational energy efficiency without sacrificing accuracy. J.J. Zhang et al. \cite{ref20} proposed adding a constant 0 bypass to the calculation unit of DLA. When the corresponding calculation unit has a hard fault, it can be bypassed as a constant 0 to mitigate the large numerical fluctuations and precision loss caused by faults. J.A. Clemente et al.\cite{ref34} proposed adding redundant connections to the Hopfield neural network accelerator and using voting logic to implement error correction. This strategy can significantly reduce fault-tolerant design overhead compared to triple modular redundancy. These works mainly enhance fault tolerance through circuit design, but also require fine-tuning of the numerical bias introduced by the circuit layer in the algorithm or model layer.

At the architecture level, Liu et al.\cite{ref3,ref12}  proposed a heterogeneous computing architecture to solve the problem of arbitrary hard faults in the DLA computation array, using a dot product computation array different from the two-dimensional pulsating array to achieve re-computation of tasks on any computing unit. C. Schorn et al. \cite{ref26} proposed dividing the DLA's computation array into highly reliable computation regions and ordinary computation regions, which are used to process fault-sensitive and fault-insensitive computing tasks, respectively. However, the distribution of fault-sensitive and insensitive computing tasks usually changes with the model or even the input, so some detailed microstructure support is still lacking. Zhen Gao et al.\cite{ref30}  proposed adding integrated learning units on top of the deep learning accelerator, which allows for the parallel computation of multiple lightweight deep learning models to tolerate hardware faults and improve inference reliability. However, it depends on the redesign of the model. E. Ozen et al. \cite{ref25} proposed adding parity check units on the deep learning computation array, utilizing algorithm-based fault tolerance (ABFT) technology to achieve real-time fault detection and correction, which is limited to low fault-rate scenarios due to the fault-tolerant and error-correcting capabilities of ABFT technology.

At the algorithm level, many works enhance the fault tolerance of deep learning models to hardware failures by mining their own fault tolerance and parameter redundancy, while ensuring accuracy through changing model parameters or even structure, without changing the DLA hardware design \cite{ref17,ref28,ref36,ref37}. These works usually rely on fault-tolerant training to change deep learning model parameters or architecture. Such model training-based fault-tolerant designs often depend on sample data and a large amount of retraining, and are sensitive to fault rates, making it difficult to fully consider during the training phase. In contrast to fault-tolerant methods that heavily rely on large amounts of application data training, some methods improve model fault tolerance by using equivalent computational methods \cite{ref6,ref24}, introducing new activation functions or numerical constraints \cite{ref32,ref33,ref35}, or using checksum mechanisms for error correction \cite{ref24}. These methods generally require only a small amount of data fine-tuning or even no application data, achieving high prediction accuracy at a low computational cost under low fault rates, which is very attractive. However, when the hardware failure rate is high, reliability drops sharply, and the cost of fault tolerance also significantly increases. Another orthogonal class of techniques to the fault tolerance methods that exploit the fault tolerance of the model itself is redundancy protection. To avoid the high computational cost of direct redundancy, many works further analyze the differences in internal fragility of deep learning models \cite{ref15,ref16,ref18,ref21,ref22,ref23,ref26}and reduce the cost of redundancy protection through differential protection methods\cite{ref26,ref27,ref28,ref38,ref39} . Since deep learning computation engines generally support layered computation patterns, the differences in fragility between model layers can be easily applied to various DLA architectures. Correspondingly, the analysis of deep learning fragility focuses on the differences between layer fragilities \cite{ref15,ref16,ref21,ref22}, while selective protection focuses on layer redundancy, which can be achieved through both spatial and temporal redundancy. Theoretically, a more granular analysis of fragility differences \cite{ref26} would be helpful for more efficient selective protection, but currently, there is still a lack of DLA architecture and circuit support.

As can be seen from the above, most fault-tolerant techniques for deep learning focus mainly on one level of circuit, architecture, or algorithm. Different techniques have significant advantages and disadvantages, and there are also various limitations in their usage scenarios. Although there are also some fault-tolerant techniques that involve the coordination of different levels of fault tolerance to improve inference accuracy or reduce fault tolerance costs.The computation unit bypass technology proposed by J.J. Zhang et al. \cite{ref20}[20] can be combined with model training to enable the model to match specific fault bypass settings, thereby reducing accuracy loss caused by bypass. However, there is still a lack of systematic cross-layer optimization design methods for fault-tolerant DLAs oriented towards soft errors. Therefore, based on the idea of cross-layer chip optimization design, this paper first proposes sensitivity analysis from two dimensions, namely, neurons dimension and the different bits of nuerons dimension, at the algorithmic layer. Based on this, selective redundancy protection technologies are explored at the architecture layer and circuit layer, respectively, to minimize fault-tolerant design costs while ensuring reliability and performance.

\section{Cross-Layer Optimization for Fault-Tolerant Deep Learning Accelerators}
To address the multi-objective design requirements for fault-tolerant DLA in various scenarios, we propose a systematic cross-layer optimization design framework that integrates the fault-tolerant advantages of circuit layer, architecture layer, and algorithm layer, while minimizing fault-tolerant cost under the premise of meeting inference accuracy and performance. We will first introduce the cross-layer optimization framework for fault-tolerant DLA, and then expand on the fault-tolerant design strategies corresponding to the algorithm layer, architecture layer, and circuit layer. Finally, we demonstrate an automated cross-layer parameter design space exploration method.
\subsection{Overall Architecture}
The overall architecture for cross-layer optimization of fault-tolerant DLA is shown in Figure 1. First, the user needs to determine the design objectives and constraints, such as performance, reliability, and resource costs. Reliability is usually measured by accuracy under fault scenarios, to some extent overlapping with the accuracy metric of the model. Resource costs mainly include additional chip area introduced by fault-tolerant design.

After the design goals and constraints are given, the framework analyzes the sensitivity of different neurons to soft faults at the algorithm level, and uses this as a basis to divide the neural computation in deep learning into important and common parts. Compared to ordinary neural computation, important neural computation is more sensitive to soft faults and leads to greater loss in model accuracy, therefore requiring stronger fault-tolerant design. On this basis, we further refine the sensitivity of neurons to soft faults from the dimension of bit width, where high bit-width errors in neurons will cause larger numerical deviations and have a greater impact on model accuracy, thus requiring relatively stronger fault-tolerant protection for high bit-width computation as well. In summary, we divide deep learning computation into important and ordinary parts from two dimensions: neurons and data bit-widths.

In addition, model quantization is a crucial step in achieving high energy efficiency in deep learning processing, enabling the use of lower data bit-widths for computation. To support fixed-point quantization, basic computation units in DLA such as multiply-accumulate units typically need to be truncated according to the quantization. Truncation not only affects precision but also directly impacts the computational logic size of the calculation unit output values in different bit-widths, as well as the selective redundancy protection cost in DLA circuitry layer, which will be detailed in our redundancy design at the circuitry layer.

Although we divide the processing of deep learning into important neuron computations and common neuron computations at the algorithmic level, it is difficult to separate and perform different fault-tolerant protections on important and common neuron computations due to the typical streaming computation used in DLA architecture design and the need for further partitioning (tiling) of deep learning models for computation on DLA through time division and reuse. Additionally, there are significant differences in the distribution of important neuron computations across different deep learning models and the distribution of important neuron computations in different partitions of the same deep learning model, further increasing the difficulty of separating important and regular neuron computations in DLA architecture. To address this issue, we extended the HyCA architecture\cite{ref3,ref12}and proposed a flexible and configurable heterogeneous DLA architecture, FlexHyCA, which uses a conventional 2D computing array to process regular neuron computations and a heterogeneous dot product processing unit (DPPU) to process a small amount of important neuron computations that require higher reliability. By loading real-time distribution information of important neurons on the 2D array and segmenting the computation of different neurons, DPPU can flexibly choose to reuse the data loaded in the 2D array cache or directly read the required data from DRAM based on the proportion of important neurons, ensuring that DPPU will not block the 2D array computation and eliminating the impact on DLA performance.

In response to the difference in fault sensitivity between important and ordinary bits of neurons in deep learning models, we designed selective bit protection circuits on the computing units of FlexHyCA. Only the important bits of ordinary neurons and the directly related logic circuits of important neurons' important bits were redundantly protected to reduce the cost of fault-tolerant design.

We found that the choices of fault-tolerant design at different levels can influence each other. For example, the proportion of important neurons at the algorithmic level directly determines the size of the DPPU in the FlexHyCA architecture, which affects the hardware resource cost. Similarly, the number of important bits in neurons also corresponds to the size of logic circuits that need to be protected, ultimately affecting the cost of bit protection circuit design. The quantization choices of the model also affect the truncation of the computational unit, thus impacting the size of the circuit that needs to be covered by fault tolerance and the associated cost. The ratio of important neurons to ordinary neurons also affects the number of important bits in each, and these design parameters interact with each other, making manual optimization of fault-tolerant DLAs very challenging and difficult to address the diverse needs of users' design goals and constraints in different scenarios. Therefore, we introduce a design space exploration mechanism based on the Bayesian algorithm, and use the mutual relationships of local parameters to prune the design space, achieving rapid automated cross-layer parameter optimization.
\begin{figure}[!t]
\centering
\includegraphics[width=3.2in]{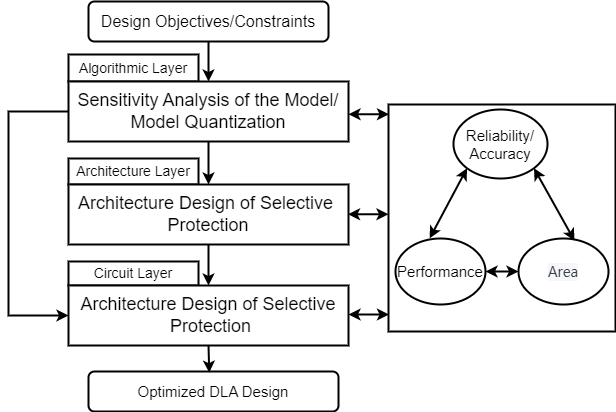}
\caption{Design framework for cross-layer optimization}
\label{fig_1}
\end{figure}

\subsection{Algorithm-level fault-tolerant design}
We first analyze the differences in the sensitivity to soft faults between models from two dimensions: neuron computation and nueron bit-width. The neural network model can be viewed as a complex function$f(x)$ where $x$ represents input data. By first-order Taylor expansion of the function, as shown in Equation (1), the first-order error $f(x+{\Delta}x)-f(x)$ caused by perturbation near the input data is proportional to the perturbation size$\Delta{x}$ and the gradient value$f'(x)$. Therefore, it can be approximated that the sensitivity of neurons is related to their gradient on data input. Neurons with larger gradients may cause larger numerical fluctuations due to perturbations, which may have a greater potential impact on the model accuracy. It can be considered that the corresponding neurons are more sensitive to faults, and we define them as important neurons. Similar methods for analyzing neuron importance have also been applied in [26][44].
\begin{equation}
    f(x+{\Delta}x)-f(x)=\Delta{x}f'(x) +O(\Delta{x})
    \label{eq1}
\end{equation}

We identify the top S\_TH\% neurons with the highest gradient values as the important neurons, and provide detailed gradient-based neuron importance analysis in Algorithm 1.

\begin{algorithm}
\caption{Gradient-based important neuron selection algorithm.}\label{alg:algorithm1}
	\KwIn{Neural network model:$M$; Dataset: $D$; Total number of neurons: $N$;
            Important neuron proportion: $S\_TH$;}
	\KwOut{set of important neurons: $S$}  
    GetImportantNeurons($M, D, N, S\_TH$)\;
       \BlankLine
       Initialize the $gradients[N]$ of each neuron\;
       \For{$i \gets 1$ \KwTo $N$}{
            \ForEach{$input \in D$}{
                $output \gets inferenceModel(M, Input)$\;
                $grad \gets backwardModel(M, Output)$\;
                $gradients[i] \gets gradients[i] + Abs(grad)$ \;
            }
        }
        Sort $gradients$ in descending order\;
        Select top $N \times S\_TH$ neurons based on $gradients$\;
        Put the selected neurons in $S$\;
        return $S$\;
\end{algorithm}

Due to the greater numerical disturbance caused by high-bit flips than by low-bit flips, we define the high $NB\_TH$ bits of each neuron as the important bit positions of ordinary neurons. Important neurons are more affected by flips and require more protected bits, so the high $IB\_TH$ bits of important neurons are defined as the important bit positions of important neurons. Due to the limited data bit width of fixed-point deep learning models and the higher importance of high bits than low bits, we always prioritize the protection of high bits for neurons of the same type. The compromise between important neurons and ordinary neurons depends on the model's precision and the cost of bit protection. We choose the settings with the lowest protection cost while meeting the precision requirements. Since the number of combinations of $IB\_TH$ and $NB\_TH$ settings is not large for fixed-point models with limited data bit width, we use a simple enumeration algorithm to determine $IB\_TH$ and $NB\_TH$, where the precision is obtained through fault injection experiments and the cost of bit protection can be obtained through logic synthesis of the units. The specific design and cost evaluation of bit protection will be described in detail in the subsequent circuit-level fault-tolerant design section, and the optimization of important bit position settings is shown in Algorithm 2.
\begin{algorithm}[t]
	\caption{Bit importance evaluation}
	\label{alg:algorithm2}
	\KwIn{Neural network model:$M$;
            Dataset: $D$; 
            Set of Important Neuron:$N$;
            Neuron data bit width:$B$;
            Model Accuracy Objective:$ACC$;}
	\KwOut{The important bits of the important neurons are the high $IB\_TH$ bits;
       The important bit in regular neurons is the high $NB\_TH$ bit. }  
         GetBitConfig($M,D,N,IB\_TH,NB\_TH,ACC$);
       \BlankLine
       
       \For{$IB\_TH \gets 1$ \KwTo $B$}{
            \For{$NB\_TH \gets 1$ \KwTo $B$}{
                $acc \gets  RunModel(M, N, D, ib\_th, nb\_th)$\;
                $cost \gets getCost(M, N, ib\_th, nb\_th)$\;
                \If{$acc > ACC$ and $cost < opt\_cost$}{
                    $opt\_ib\_th, opt\_nb\_th \gets ib\_th, nb\_th$\;
                    $opt\_cost \gets cost$\;
                }
            }
        }
        return $opt\_ib\_th,opt\_nb\_th$;
\end{algorithm}

Quantization is a crucial step in deep learning model quantization, which not only affects the accuracy of the model but also affects the high-level logic in DLA computing units. For quantized DLAs, the output data bit-width of the computing unit is usually larger. For example, in an 8-bit DLA, the output data bit-width of the computing unit is at least 16 bits. To prevent overflow, many DLA designs choose settings larger than 16 bits, even up to 32 bits. Since the calculations of subsequent layers in the model are still 8 bits, the output of the computing unit needs to be truncated during the calculation process according to the model's quantization. This means that the data bits outside of the truncation have little impact on the model inference, and the importance of the logic circuits directly related to these bits is relatively low. Similarly, the importance of the logic circuits corresponding to the ordinary bit positions in the output data is also relatively low. However, the quantization selection of general deep learning models has no limitations. Different deep learning models and different layers of the same model may have different quantization selections, which leads to different important logic circuits in the computing unit. In order to ensure the worst-case scenario in DLA design, the weight of the important logic circuits is significant. Figure 2 shows the positions of the important bit positions of the accumulators and multipliers corresponding to different quantization selections, as well as the direct calculation logic areas. Assuming an input data bit width of 8 bits, a multiplier output of 16 bits, and an accumulator output of 24 bits, the top 2 bits of the output data are important bit positions. When the output data location intercepted by the accumulator is from the 2nd to the 9th bit, the top 2 bits of the output data are important bit positions. The corresponding direct calculation logic corresponds to the blue area in Figure 2(a), where the important calculation logic of the multiplier includes a column of 8 (1-bit sums) and a column of 7 (1-bit sums). Without quantization constraints, the bit positions 6 to 15 of the multiplier output may all be important, and the circuit area that needs protection corresponds to the entire red dotted line area, which will inevitably lead to high redundancy costs. We define the lowest bit of the quantized truncated data as Q\_scale. When we set Q\_scale=5, the quantization is limited so that the range of the truncated data bits is reduced to the 5th to 24th bits. The range of the top 2 bits of the multiplier is from the 11th to the 15th bit. The calculation logic area that needs to be protected in the multiplier can be significantly reduced, as shown by the red dotted line area in Figure 2(b). When we perform three-mode redundancy protection on the important calculation logic, quantization constraints can significantly reduce the cost of redundancy protection.
\begin{figure}[htbp]
  \centering
  \captionsetup[subfloat]{labelfont=footnotesize,textfont=footnotesize}
  \subfloat[Data with unquantized limits and accumulator truncation can be located at any contiguous position of 24-bit data]{
    \includegraphics[width=3in]{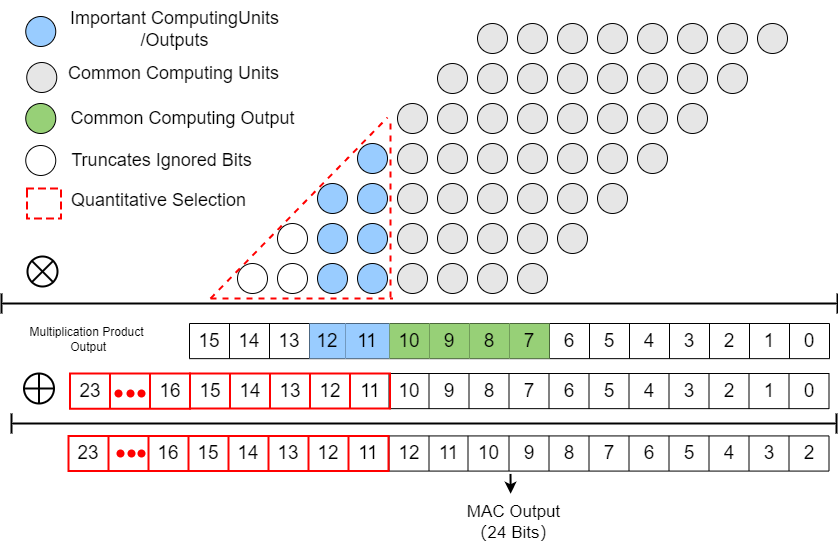}
  }
  \hfill
  \subfloat[There are quantization constraints, and the data truncated by the accumulator can be located at any continuous position between the 5th and 23rd bits]{
    \includegraphics[width=3in]{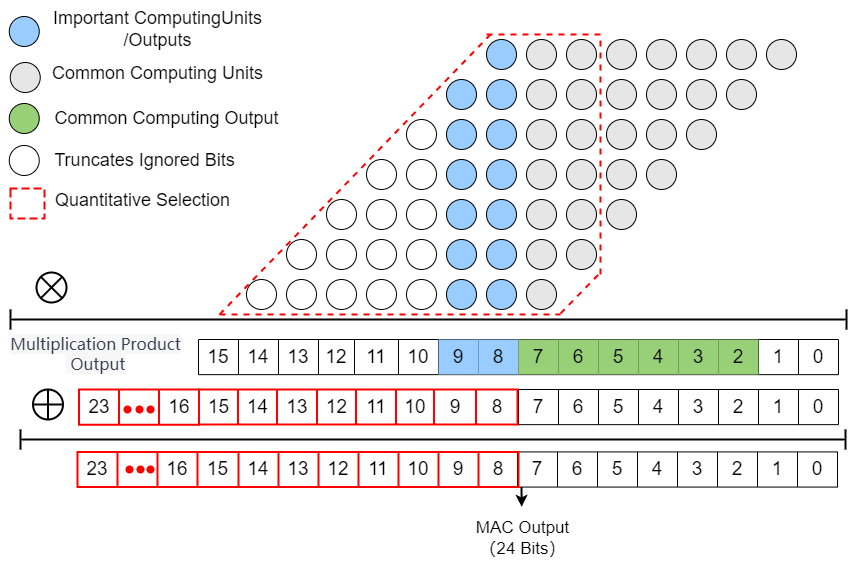}
  }
\caption{The range of the important bits of the accumulators and multipliers in DLA, as well as the directly associated computational logic, vary under different quantization constraints.}
\end{figure}

\subsection{Architecture-level fault-tolerant design}
In order to support selective fault-tolerant protection, the core issue faced by the design of the DLA architecture is how to differentiate between important neuron computations and normal neuron computations on top of the basic DLA architecture, and provide stronger fault-tolerant protection for a small number of important neurons. Specifically, the architecture design must also be able to tolerate the variability in the distribution of important neurons. The variability in the distribution of important neurons is mainly reflected in three aspects: (1) the distribution of important neurons in different deep learning models varies; (2) the distribution of important neurons in different layers of the same deep learning model is also not the same; (3) the convolution of the same deep learning model in the same layer often needs to be partitioned and mapped to the computational array of DLA in a time-shared manner due to the limited resources of DLA, and the distribution of important neurons on different task slices will still vary.

To achieve this, we utilize HyCA, proposed by Liu et al. [3][12], for arbitrary position computing unit fault tolerance in DLA, as the basis infrastructure to separate the computation of important neurons from ordinary ones. Since most of the neurons are classified as ordinary, we use the two-dimensional computation array of HyCA to process ordinary neuron computation, in order to better reuse data and weights. For a small number of sparsely distributed important neurons, we dynamically load them into the heterogeneous dot product processing unit (DPPU) for processing, fully utilizing the internal parallelism of neuron computation. Since the DPPU of HyCA needs to reuse data in the dedicated cache of the two-dimensional array, when the proportion of important neuron computation that needs to be protected is too high, the DPPU will block the computation of the two-dimensional array. On the other hand, the proportion of important neuron computation varies greatly, and setting the DPPU size for worst-case scenarios will introduce significant area overhead. To address this issue, we have added a more flexible data loading mechanism to the DPPU on top of HyCA. When the proportion of important neuron computation is too high, the required data is loaded directly from DRAM, avoiding the performance degradation caused by DPPU blocking the two-dimensional array computation, while only adding a small amount of I/O. When the distribution of important neurons is relatively uniform, FlexHyCA is compatible with the original HyCA design, prioritizing the reuse of data in the two-dimensional array cache to avoid additional I/O. This new FlexHyCA architecture can better adapt to the problem of differences in the distribution of important neurons, as shown in Figure 3.
\begin{figure}[!t]
\centering
\includegraphics[width=3in]{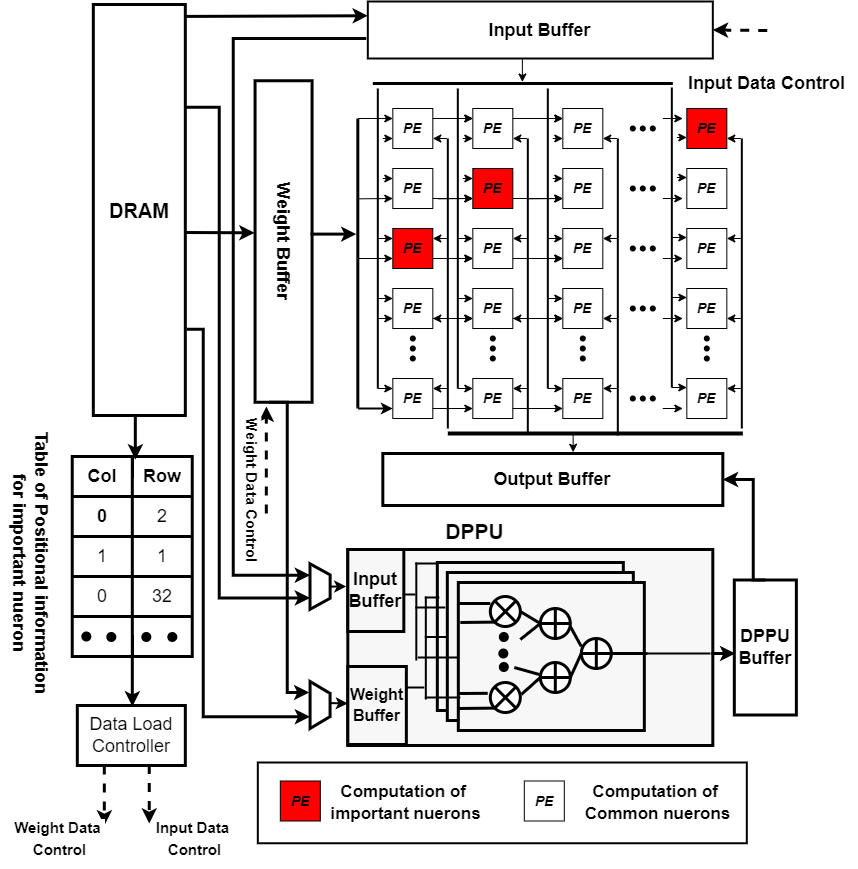}
\caption{FlexHyCA Architecture}
\label{fig_3}
\end{figure}

\subsection{Circuit-level fault-tolerant design}
To support selective fault-tolerant protection, we propose a configurable bit protection design for the basic multiply-add units in DLA based on the differences in importance of individual bit positions in the neurons. We provide triple modular redundancy (TMR) protection for logical calculations on important bit positions of neurons to avoid significant computation biases caused by soft faults. Since the core of the multiply-add unit is the multiplier section, we illustrate a configurable bit-protected multiplier based on an 8-bit basic multiplier as an example. As shown in Figure 4, the multiplier output data width is 16 bits. When the 8-bit data truncated by the accumulator corresponds to the [m:n] bits of the multiplier output, the most significant bits of the multiplier correspond to the [m:m+s-1] bits. Therefore, we consider the computation logic from column m-s+1 to column m as the protected area. Since the position of m varies with the choice of quantization, we need to protect the entire red area indicated in Figure 4. However, for a specific choice of quantization, we only use a continuous s columns in the red dashed line area for computation. To reduce overhead, we only need to protect the two largest blue columns in the area. For different choices of quantization, we need to change the Mux selection to enable the redundancy units to protect the corresponding columns. However, the computation tasks on the left of the red dashed line area are much smaller than the largest two computation tasks. When a simple Mux replacement strategy is used, there will be significant waste of redundancy array when the left two columns are selected for computation in quantization, and the redundancy calculation units correspond to a large fan-out of signals, introducing additional delay and area. To solve this problem, we merge adjacent calculation units on the left, and if any column of the merged units is selected as an important computation unit, the merged units will be protected, further improving the utilization of redundancy units and reducing the fan-out of signals. When s=1 and Q\_scale=2, as shown in Figure 4, the maximum fan-out of the redundancy calculation units for the leftmost 3 columns is reduced from 6 to 4. When quantization results in the most important column being in the left area, redundancy calculation can also be used to protect more computation columns. Although Figure 4 only shows the effect of bit protection for a conventional shift multiplier, the bit protection strategy can also be applied to other multiplier structures. Figure 4(b) illustrates an example of applying the bit protection strategy to a Wallace tree multiplier.
\begin{figure}[htbp]
   \centering
   \captionsetup[subfloat]{labelfont=footnotesize,textfont=footnotesize}
   \subfloat[The schematic diagram of the shift multiplier bit protection]{
    \includegraphics[width=3in]{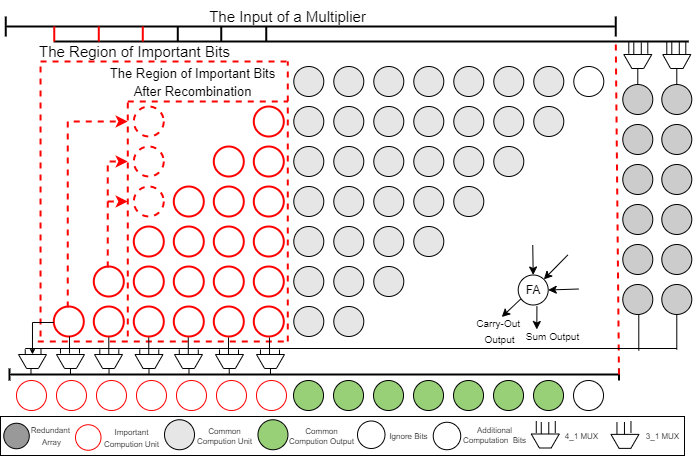}
  }
  \hfill
  \subfloat[The schematic diagram of the Wallace tree multiplier bit protection]{
    \includegraphics[width=3in]{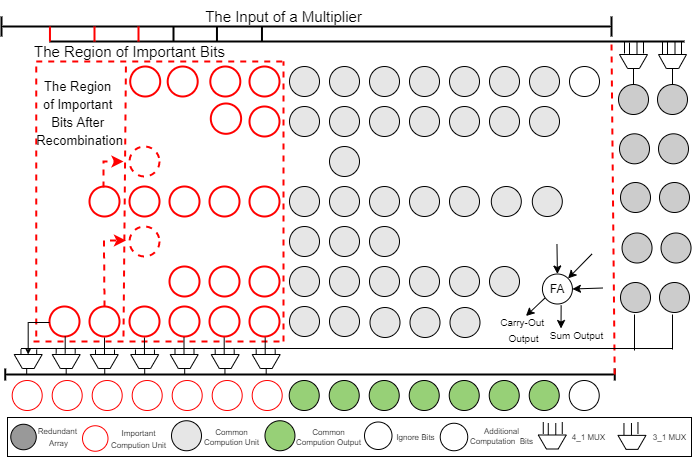}
  }
  \hfill
\caption{Configurable bit-protecting multiplier schematic (Q\_scale=2).}
\end{figure}
\subsection{Cross-layer spatial exploration design}
Due to the interconnected of design parameters across different layers, optimizing each layer separately is difficult to achieve the optimal design goal. Therefore, we constructed a unified cross-layer optimization space to collaboratively optimize the design parameters of different layers. We specified the accuracy requirement and performance loss of the model under a designated fault rate as the user's design constraints, while minimizing the additional chip area introduced by redundancy protection. The design parameters of the algorithm layer mainly include the percentage of important neurons in the model (S\_TH\%), the selection strategy of important neurons (S\_policy), the number of important bit positions for important neurons (IB\_TH), the number of important bit positions for ordinary neurons (NB\_TH), and the quantization truncation parameter (Q\_scale). The design parameters of the architecture layer mainly include the two-dimensional array size (Array\_size) and dot product array size (Dot\_size) of FlexHyCA, and whether the heterogeneous array data is reused or not (Data\_reuse). The design parameters of the circuit layer include the implementation strategy of bit redundancy for computing units (PE\_policy), such as direct redundancy or configurable redundancy.

The cross-layer parameter design space for fault-tolerant DLA is large, and many parameters will affect different layers. It is difficult to achieve complex design goals manually, and the design efficiency is relatively low. To solve the problem of cross-layer parameter optimization, we first formalized the cross-layer parameter optimization problem. As shown in Equation 2, all design parameters are uniformly represented by a vector V, and each component corresponds to the design parameters of different layers. We use performance, reliability, and accuracy as constraints, set the typical fault rate as the reliability and accuracy indicator for model inference, and minimize the chip area introduced by redundant design as the optimization goal.

To solve the problem of cross-layer parameter optimization for fault-tolerant DLA, we use the Bayesian optimization algorithm as the basic space exploration method and implement it as shown in Algorithm 3. We also added a limit on the maximum iteration number (ITER\_MAX\_STEP). Each iteration needs to evaluate the corresponding configuration's reliability, accuracy, performance, and redundant chip area. We use Scale-Sim [19] to evaluate performance, Synopsys Design Compiler to evaluate the area of bit-protected computing units. Due to the repetition of many bit-protected options in the design space, we pre-evaluated the cost of possible bit-protected designs and constructed an area cost table. In the Bayesian optimization process, different configurations of FlexHyCA area can be quickly obtained by looking up the table. We use our self-developed fault injection simulator to evaluate the model's accuracy under different fault rates. To speed up the search of the design space, we also use empirical parameters to prune some parameters in the design space in advance. For example, more bit protection can improve the model's fault tolerance, and under the same fault rate, the corresponding accuracy is higher, but it will also introduce higher redundant protection costs. Essentially, the bit protection parameter is monotonically increasing with both accuracy and chip redundancy. Similarly, with other settings unchanged, the proportion of important neurons and the accuracy and redundancy costs are also monotonically increasing. This empirical information can be used to quickly prune design choices that do not meet the constraints. For example, if the bit protection settings violate the accuracy or reliability requirements, fewer bit protections will also violate the accuracy or reliability requirements, so such design parameters can be skipped without evaluation.

\begin{align*}\label{eq2}\tag{2} 
    \makecell[c]{V=S\_TH,IN\_TH,NB\_TH,Q\_scale,\\S\_policy,Array\_size...} \\
    \makecell[c]{arg\ \underset{v\epsilon V }{min} \ Area \\
    s.t \qquad ACC_{high}\ge  0.97ACC_{0} \\
    ACC_{low}\ge  0.95ACC_{0} \\
    Perf\le 1.10Perf_{0} \\
    Bandwidth \le 1.10Bandwidth_{0} \\}
\end{align*}

\begin{algorithm}[t]
	\caption{Gradient-based important neuron selection algorithm.}
	\label{alg:algorithm3}
	\KwIn{Design Parameter Space for Cross-Layer Design:$V$; Design Constraints: $R$;}
	\KwOut{Optimal Parameter Selection:$v$}  
	\BlankLine
        BayesDesignOpt($V, R$);
        $v \gets sample($V$), step \gets 0$\;
       \While{$step < ITER\_MAX\_STEP$}{
		 $Area, Acc, Perf \gets  getDesignVal(v)$\;
             $v \gets bayesOptStep(v, V, Area), step \gets step+1$\;
            \If{not meetRestriction($Acc, Perf, Bandwidth, R$)}{
                 $v \gets bayesOptPurning(v, V)$\;
            }
        }
        
return $V$
\end{algorithm}

\section{Experiment}
\subsection{Experiment Setup}
This section mainly introduces the experimental settings from the perspectives of optimization objectives and experimental constraints, dataset and model testing benchmarks, hardware implementation configuration, and software simulation configuration.

\noindent\textbf{User optimization objectives and design constraints:} We refer to prior works such as \cite{ref10,ref22} and use the bit error rate (BER) to describe the probability of soft errors, although it represents to some extent the probability of single particle flips occurring in each storage unit containing a cache and a register on-chip, it is actually an abstract fault rate for deep learning applications that differs from specific target hardware. For two scenarios, fault rate I (BER = 1E-4) and fault rate II (BER = 2E-4), we set two different reliability/accuracy constraints. Compared to a fault-free deep learning model, the accuracy loss is less than 3\% under fault rate I and less than 5\% under fault rate II, with a performance loss and bandwidth loss both less than 10\%. The chip area cost is minimized under the premise of meeting accuracy, performance, and bandwidth constraints.

\noindent\textbf{Dataset and baseline models:} In our experiments, we employed ImageNet as the dataset and VGG16 and Resnet50 as typical deep learning models for benchmark testing. Both models were quantized using 8-bit integer quantization, resulting in quantized model accuracies of 72.95\% and 75.96\%, respectively.

\noindent\textbf{Hardware experiment configuration:} The 2D computing array in the FlexHyCA base design is fixed at 32x32, with a weight cache of 512KB and a data cache of 256KB. The basic computing unit uses a Wallace tree to implement multiplication, and the accumulator data width is 24 bits. The parameters of the FlexHyCA 2D array are fixed and do not change based on user constraints and target design, while the cache and computing array size of the 1D array are determined by the optimization tool. We modeled FlexHyCA in Verilog and obtained chip area under different configurations using Synopsys DC tools under the TSMC 65nm process.

\noindent\textbf{Software experiment configuration:} We implemented performance simulation for FlexHyCA architecture based on SCALE-Sim \cite{ref19} to evaluate the performance of deep learning under different hardware settings. We implemented fault injection simulation in PyTorch to evaluate the accuracy loss of deep learning models under different fault rates. To facilitate the implementation of fault injection in PyTorch, we referred to previous works such as \cite{ref10,ref22,ref26,ref28} and mainly performed random bit flip fault injection on neurons and weights.

\noindent\textbf{Comparison settings of fault-tolerant DLAs:} To verify the proposed cross-layer optimized fault-tolerant design method for DLAs, we compared the basic DLA design (Base), circuit-level selective triple modular redundancy design (TMR-CRT1, TMR-CRT2, TMR-CRT3), architecture-level selective triple modular redundancy design (TMR-ARCH), algorithm-level selective triple modular redundancy design (TMR-ALG), and the cross-layer redundancy design proposed in this paper (TMR-CL) in terms of hardware reliability (accuracy), hardware resource overhead, and performance. The circuit-level triple modular redundancy design mainly protects all basic computing units of DLA, as in [31], by only protecting the high bit-width operation parts. We selected the high-1-bit triple modular redundancy (TMR-CRT1), high-2-bit triple modular redundancy (TMR-CRT2), and high-3-bit triple modular redundancy design (TMR-CRT3). The architecture-level selective triple modular redundancy (TMR-ARCH) mainly protects the fault-sensitive layers based on the sensitivity difference between layers. Triple modular redundancy is implemented through spatial multiplexing and real-time voting selection. The basic DLA is divided into three equal parts, and voting logic is added. For the layers that are relatively less sensitive to faults, normal calculations are performed. The algorithm-level selective triple modular redundancy (TMR-ARG) does not require changes to the DLA architecture and repeats the execution of the fault-sensitive deep learning layers through time-division multiplexing. It can be easily implemented on deep learning accelerators or general-purpose computing platforms \cite{ref28}. The algorithm-level selective triple modular redundancy is essentially similar to the architecture-level selective triple modular redundancy, but the architecture-level selective triple modular redundancy can also be used to protect critical control paths \cite{ref15}. However, since this experiment mainly focuses on protecting data paths, the evaluation of architecture-level selective redundancy is not sufficient. For the cross-layer optimized design proposed in this paper, the design parameters are determined by the design space exploration process, and the exploration process and optimized parameters will be described in detail in Section 5.6.

The differential sensitivity of different layers in deep learning is a crucial basis for many selective redundant protection strategies. Firstly, we obtained the sensitivity comparison of different layers in VGG16 and ResNet50 through fault injection experiments. At a specified fault rate, we took the improvement in accuracy of a single layer deep learning model under complete protection relative to the situation where the entire model was unprotected as the sensitivity of that layer. To simplify the analysis, we treated a block in ResNet as a whole for analysis. Figure 5 shows the sensitivity data of different layers in VGG16 and ResNet50 under fault rate I and fault rate II, indicating significant differences in sensitivity among different layers, with the highest sensitivity layer differing from the lowest layer by more than 10 percentage points. Based on the sensitivity differences between model layers, we further analyzed the accuracy improvement brought by layer-by-layer protection according to sensitivity differences. The accuracy improvement curve is shown in Figure 6, from which we can see that the speed of accuracy improvement is initially rapid and then becomes relatively slow, further demonstrating the effectiveness of selective fault tolerance based on sensitivity analysis. Additionally, we can also determine the set of model layers that need to be protected at least based on user reliability/accuracy requirements according to this figure. In the experiment, both TMR-ARCH and TMR-ALG decided the set of deep learning model layers or blocks that needed to be protected primarily based on the aforementioned sensitivity.
\begin{figure}[htbp]
   \centering
   \captionsetup[subfloat]{labelfont=footnotesize,textfont=footnotesize}
   \subfloat[VGG16]{
    \includegraphics[width=3in]{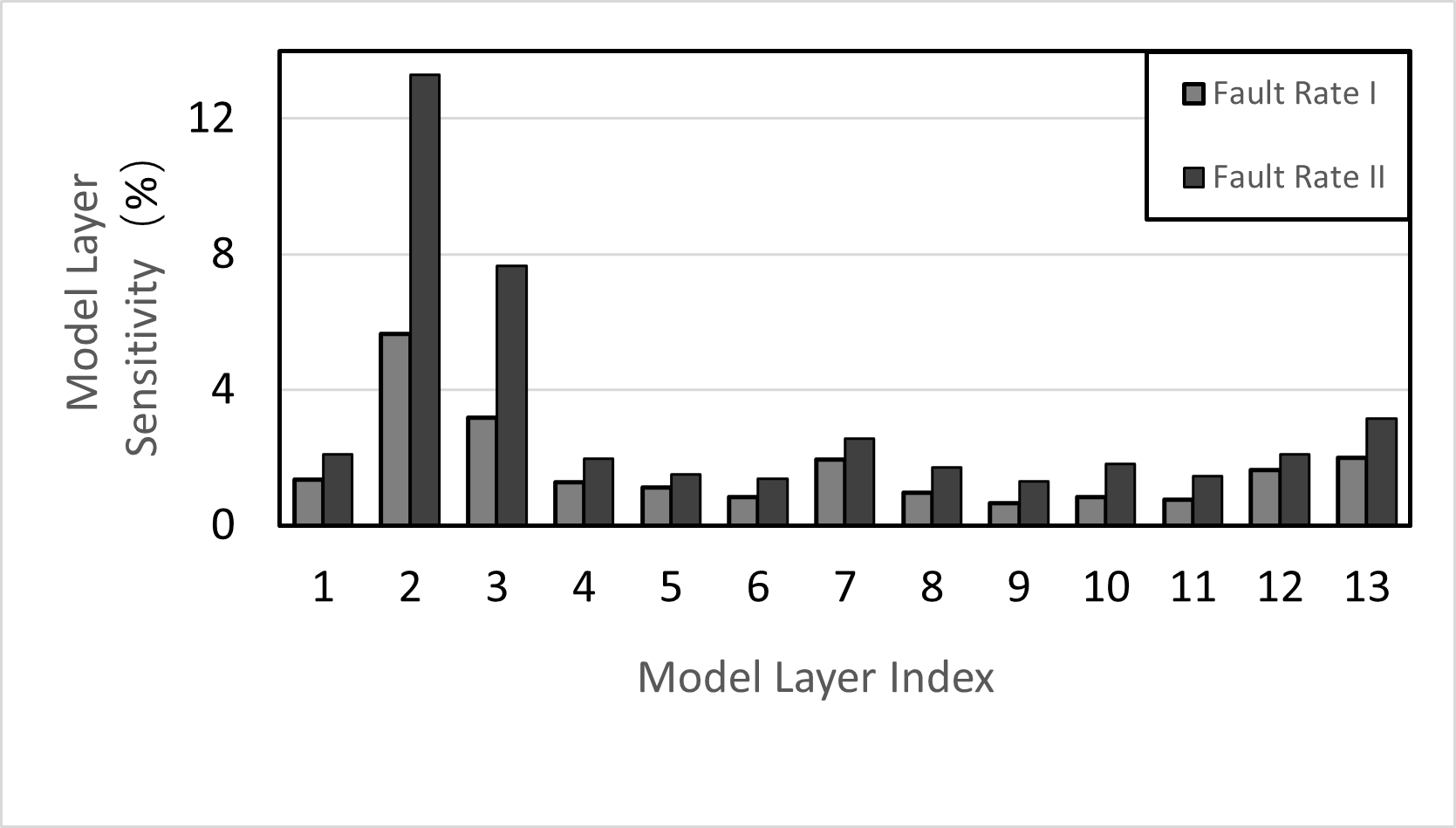}
  }
  \hfill
  \subfloat[Resnet50]{
    \includegraphics[width=3in]{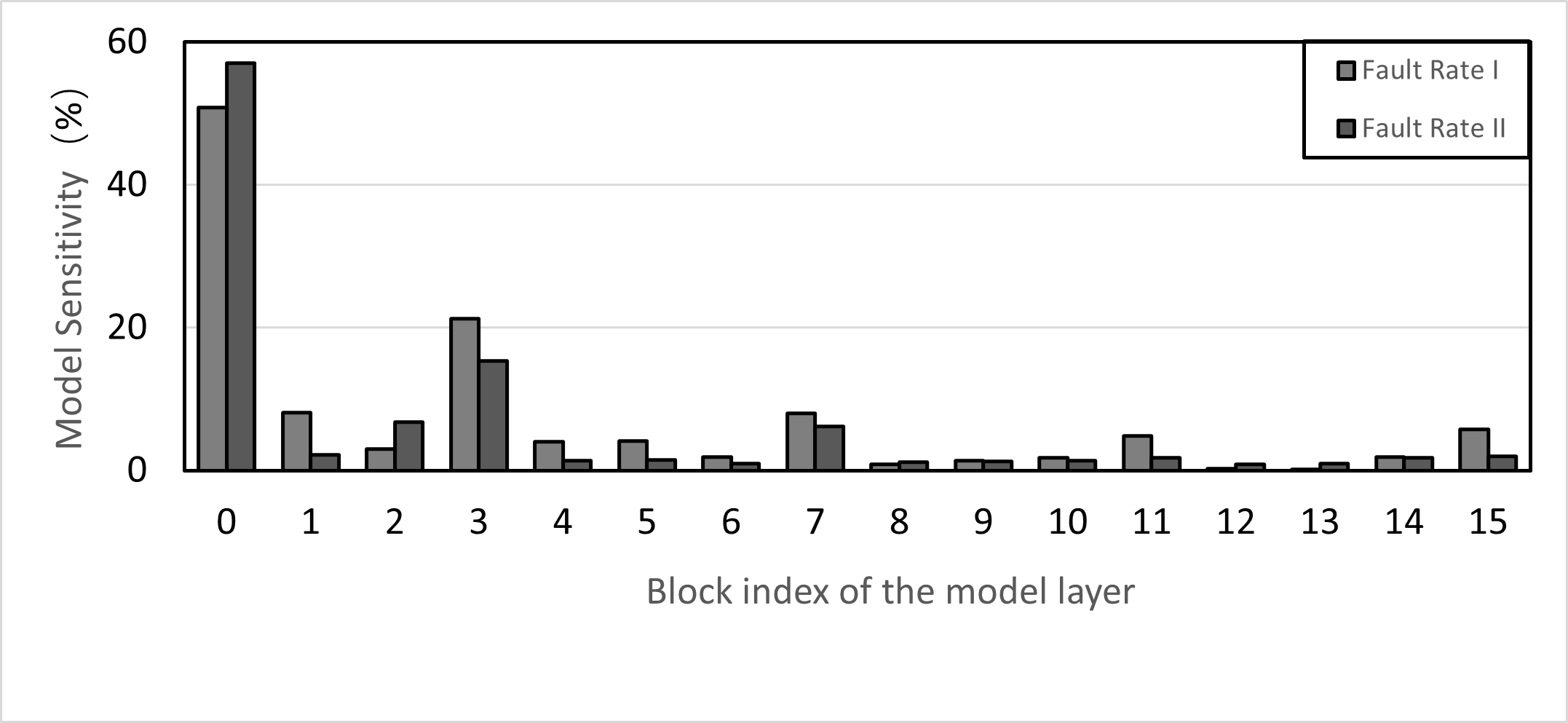}
  }
  \hfill
\caption{The sensitivity of different layers/blocks of a deep learning model under different fault rates.}
\end{figure}

\begin{figure}[htbp]
   \centering
   \captionsetup[subfloat]{labelfont=footnotesize,textfont=footnotesize}
   \subfloat[VGG16]{
    \includegraphics[width=3in]{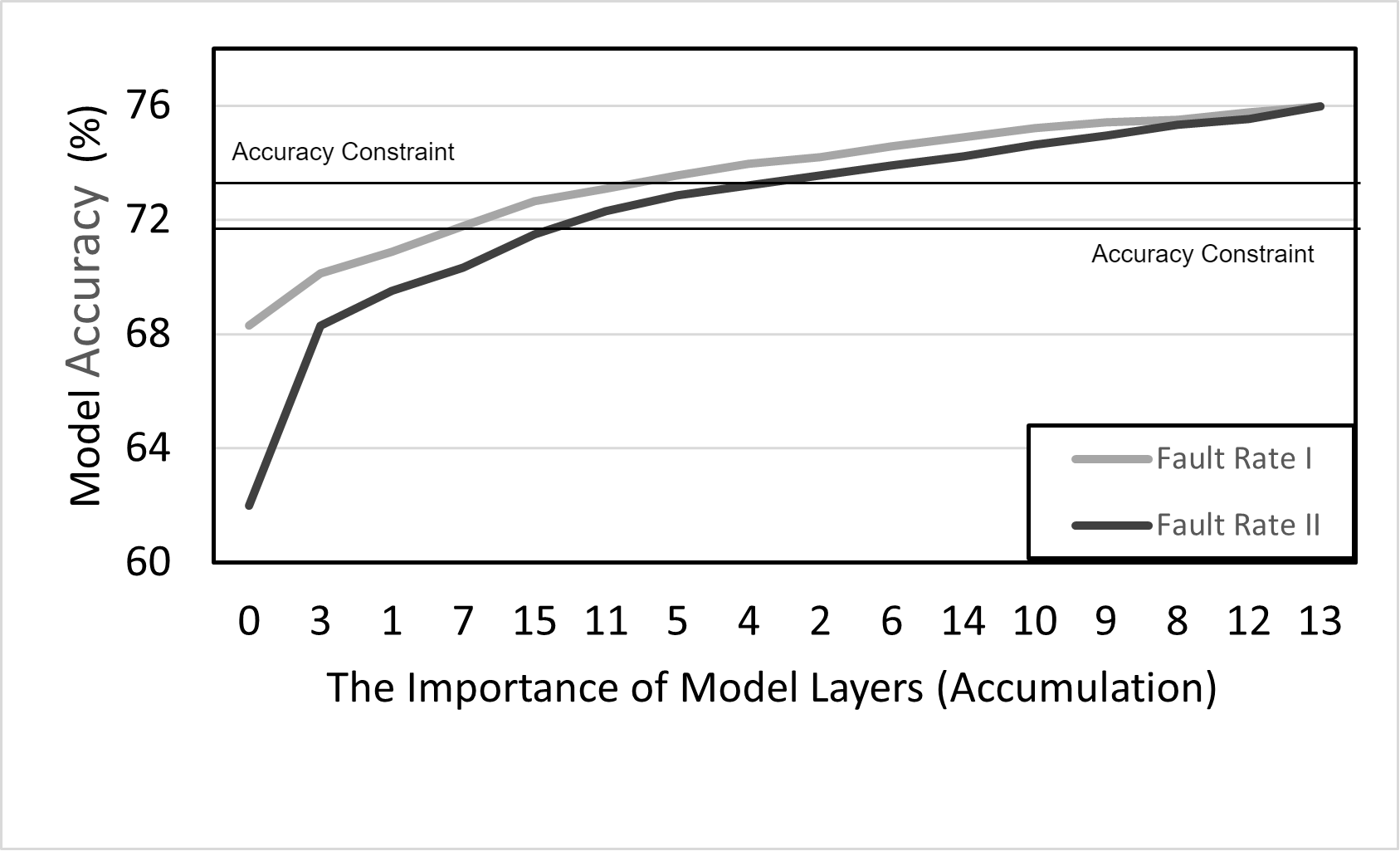}
  }
  \hfill
  \subfloat[Resnet50]{
    \includegraphics[width=3in]{figure/figure6_b.png}
  }
  \hfill
\caption{Changes in model accuracy due to sensitivity-based layer/block protection at different fault rates.}
\end{figure}
\subsection{Overall Analysis of the Experiment}
In order to optimize the objectives and constraints of the experiment, we selectively applied protection at three levels: circuit, architecture, and algorithm. We then compared the results from three perspectives: reliability/accuracy, performance, and resource consumption with both the basic design and the cross-layer design proposed in this paper.

\noindent\textbf{Reliability/Accuracy:} As shown in Figure 7, we compared the accuracy of typical fault-tolerant DLA designs under fault rates I and II in our experimental settings. Overall, different fault-tolerant strategies can meet the accuracy requirements for use. However, designs such as TMR-ARCH, TMR-ALG, and TMR-CL are relatively flexible, allowing for the discovery of settings that precisely meet the user's accuracy constraints. In contrast, the granularity of TMR-CRT is relatively coarse, and the accuracy differences are significant. TMR-CRT1 was unable to meet the accuracy constraints under fault rate II, while TMR-CRT2 and TMR-CRT3 exceeded the user's accuracy requirements.
\begin{figure*}[t]
   \centering
   \captionsetup[subfloat]{labelfont=footnotesize,textfont=footnotesize}
   \subfloat[Accuracy of VGG16 in failure rate I]{
    \includegraphics[height=0.25\textwidth]{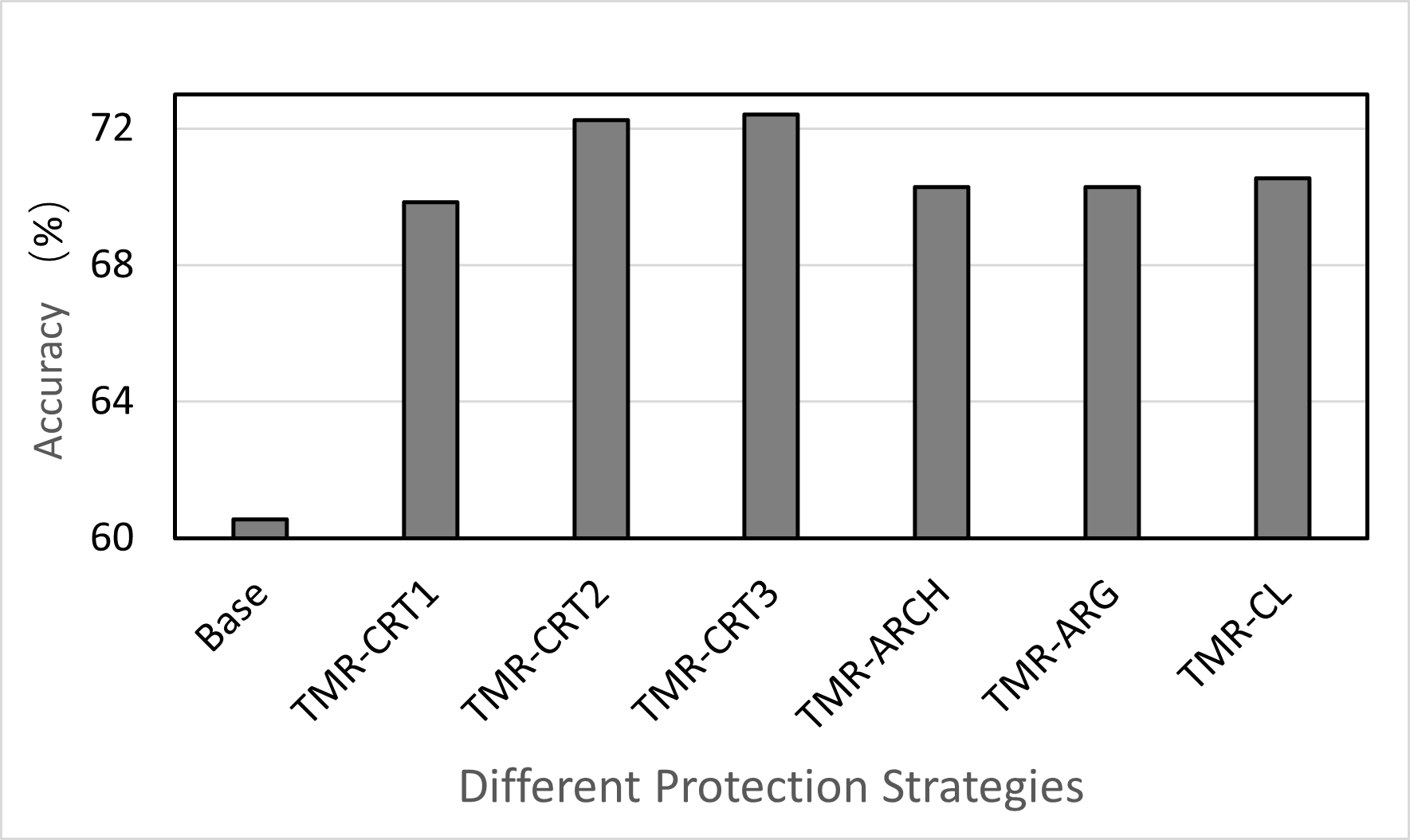}
  }
  \hfill
  \subfloat[Accuracy of VGG16 in failure rate II]{
    \includegraphics[height=0.25\textwidth]{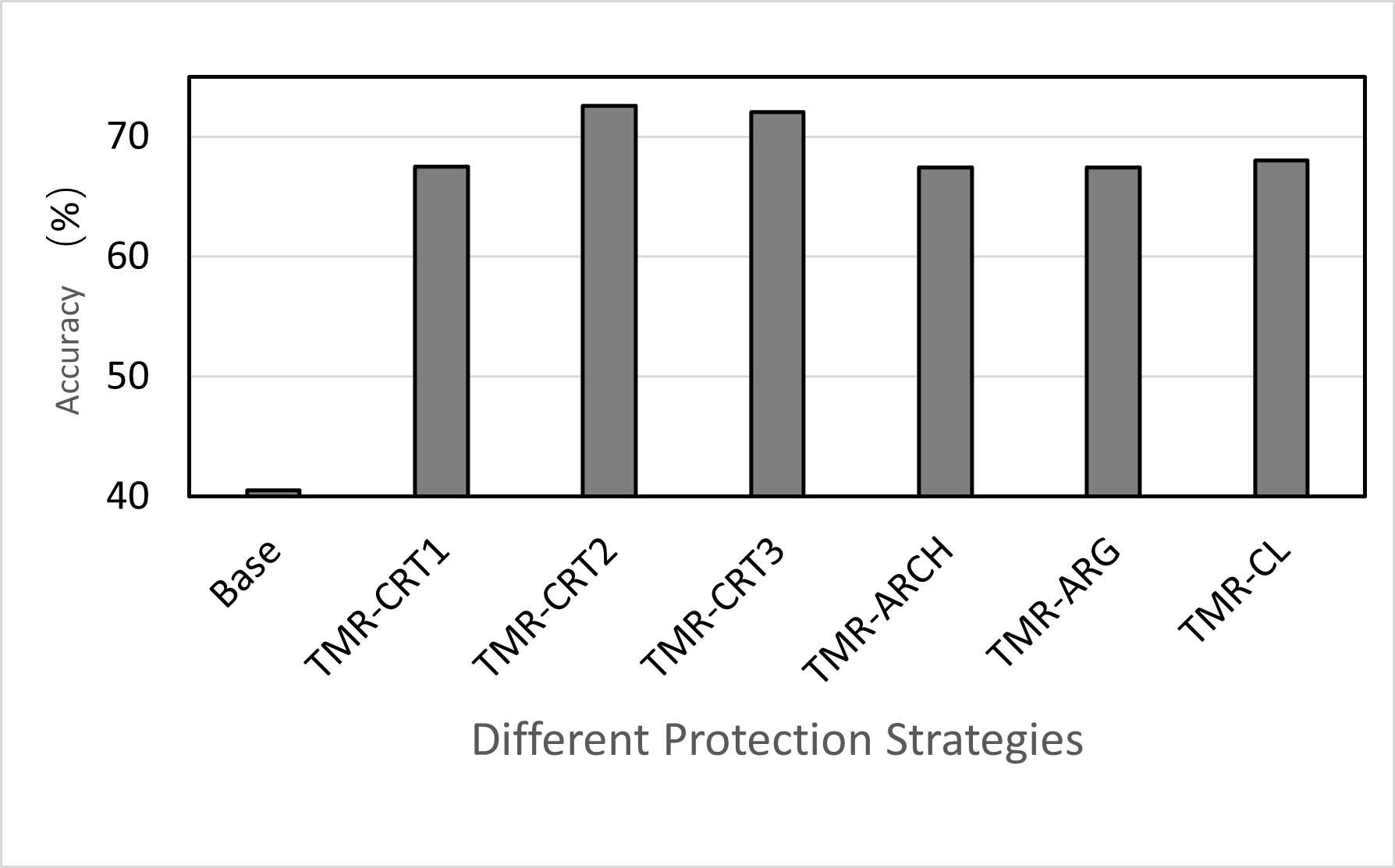}
  }
  \hfill
   \subfloat[Accuracy of ResNet50 in failure rate I]{
    \includegraphics[height=0.25\textwidth]{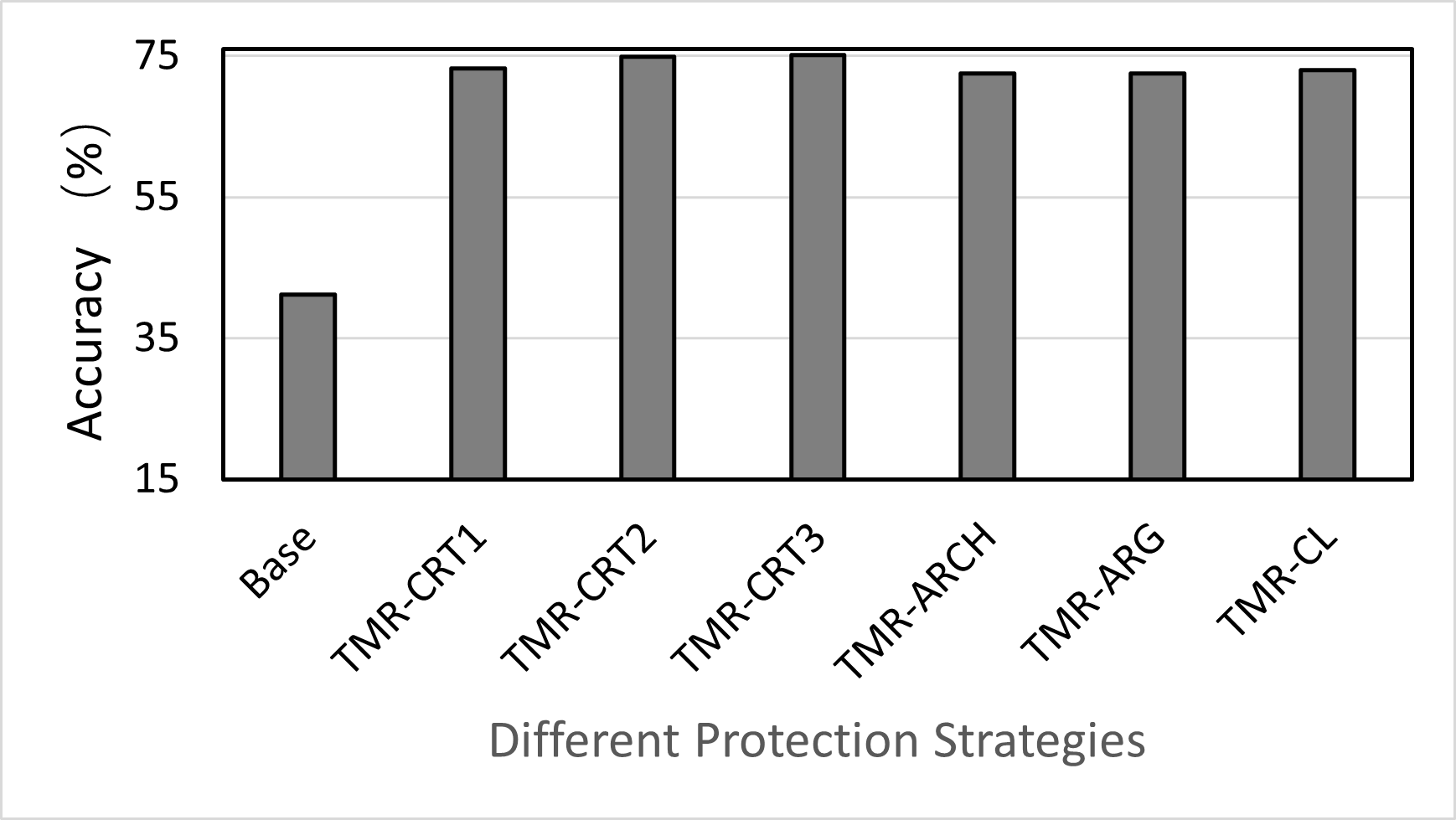}
  }
  \hfill
   \subfloat[Accuracy of ResNet50 in failure rate II]{
    \includegraphics[height=0.25\textwidth]{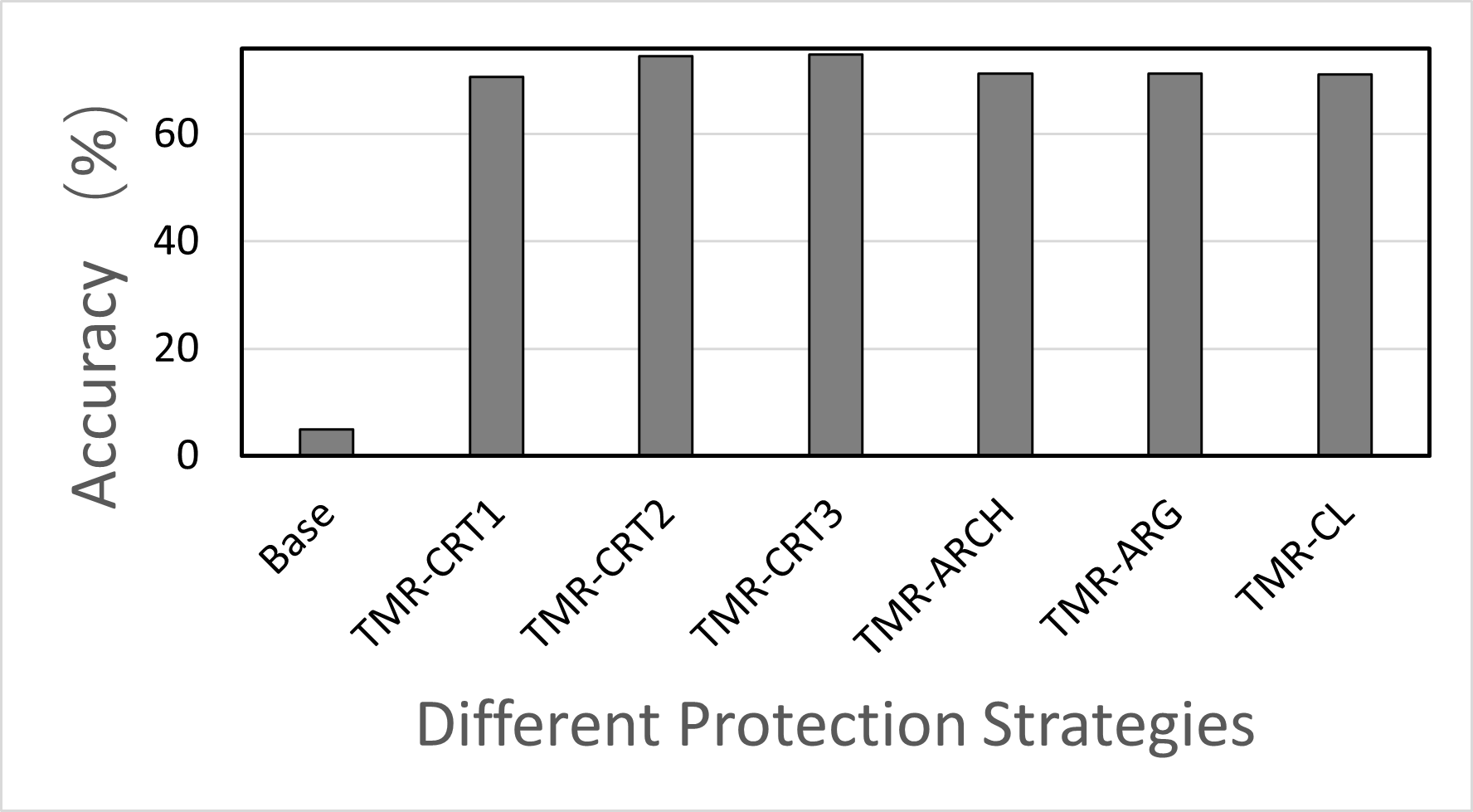}
  }
  \hfill
\caption{Model accuracy optimized by different fault-tolerant DLA design strategies}
\end{figure*}

\noindent\textbf{Performance:} The performance data is shown in Figure 8. TMR-CRT does not alter the architecture of DLA or upper-layer software, and does not result in any performance loss. TMR-CL adopts a strategy of distributing important neurons evenly across layers, so that the recalculations of important neurons are matched with the size of the DPPU, and the differences between blocks can be solved with a small amount of IO cost. Essentially, the high proportion of important neurons in one block can be shared with the low proportion of important neurons in another block, so the overall impact on performance can be negligible. In contrast, at the architecture and algorithm levels, selective redundancy is achieved through spatial and temporal multiplexing, respectively, with little or no hardware cost, but theoretically, the sensitivity of the layers where redundancy is introduced will reduce performance to one-third of the original. The sensitive deep learning model layers have a relatively high proportion, which ultimately leads to an increase in execution time of nearly double, making it impossible to meet design constraints.
\begin{figure}[htbp]
   \centering
   \captionsetup[subfloat]{labelfont=footnotesize,textfont=footnotesize}
   \subfloat[VGG16]{
    \includegraphics[width=3in]{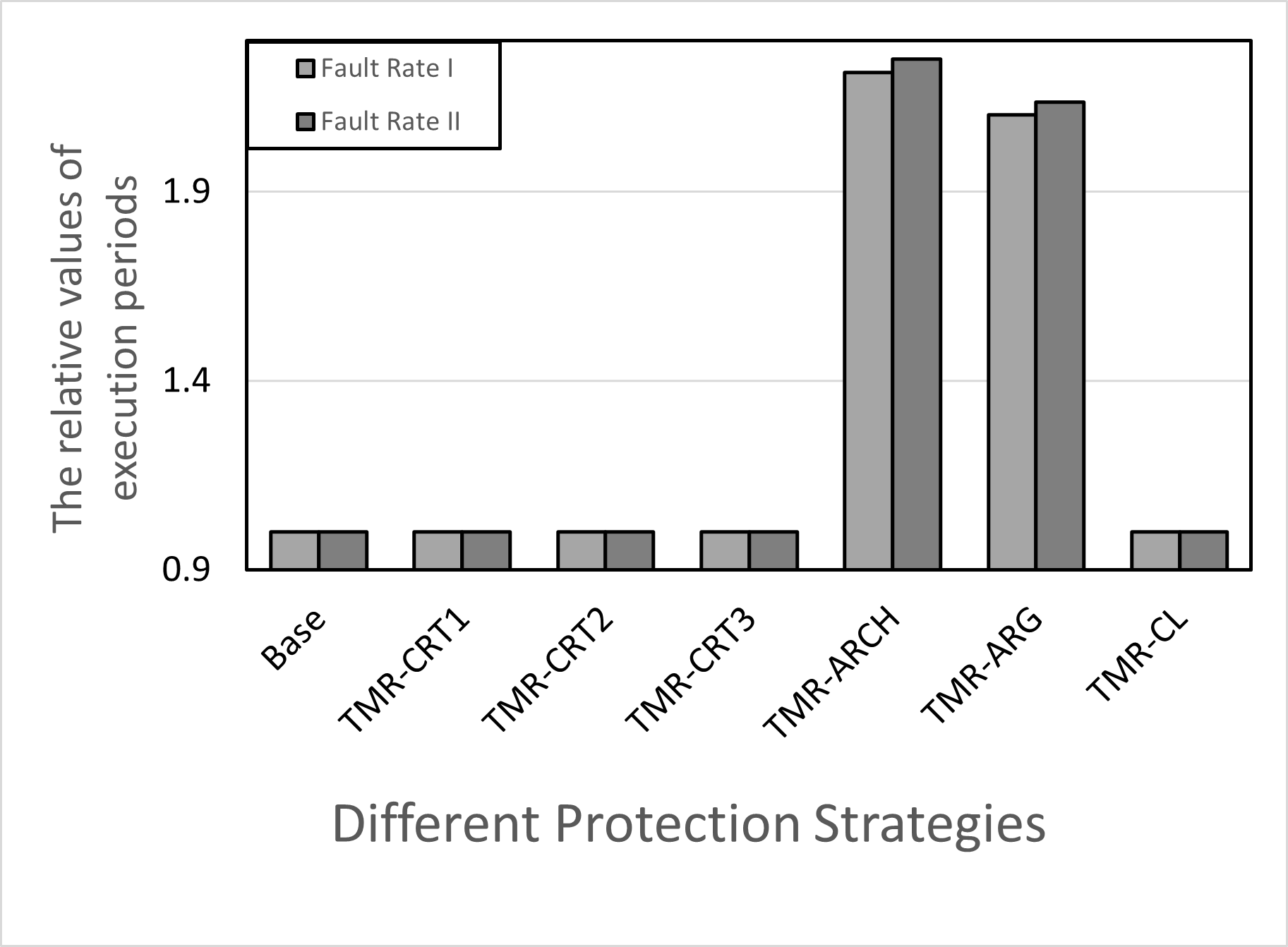}
  }
  \hfill
  \subfloat[Resnet50]{
    \includegraphics[width=3in]{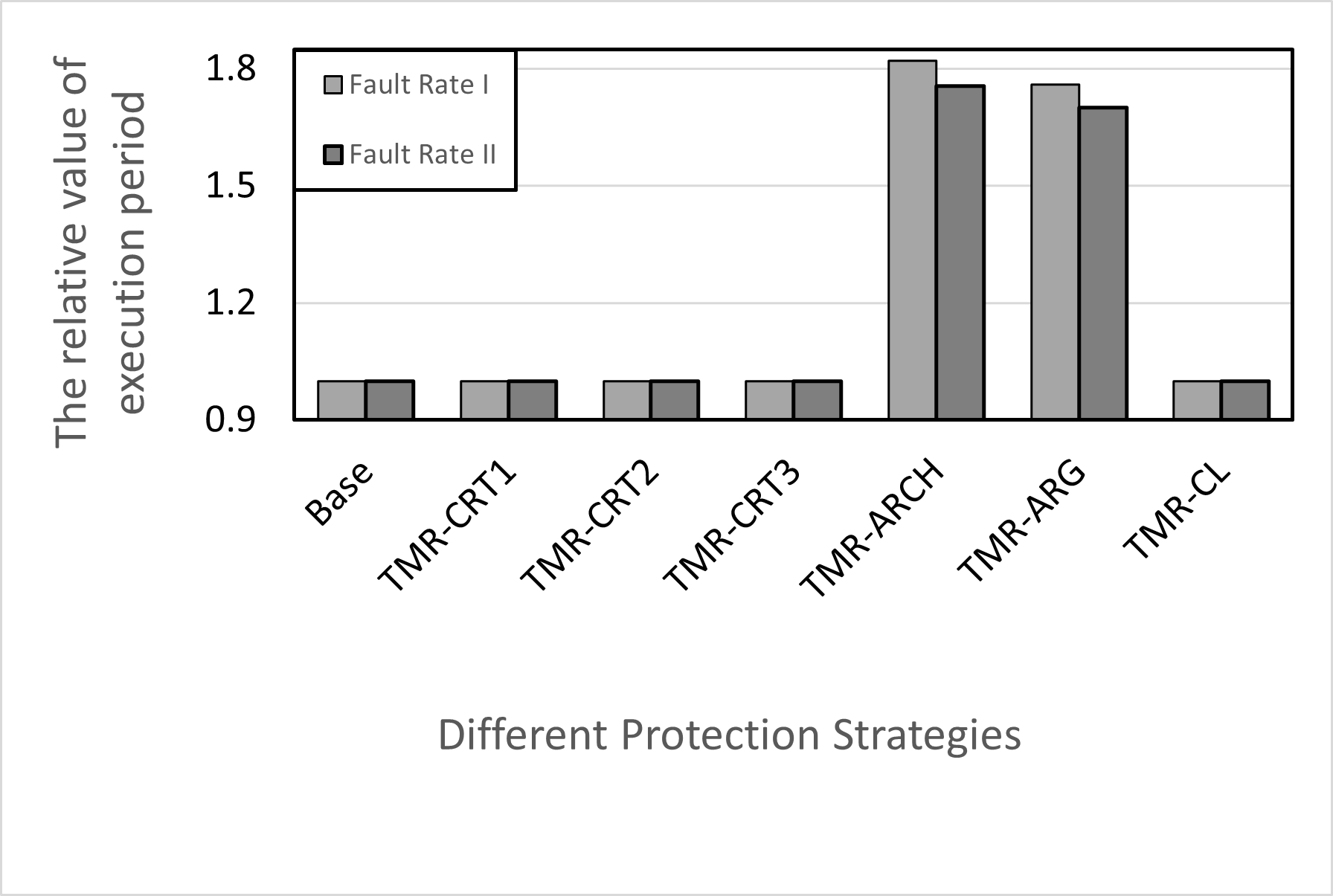}
  }
  \hfill
\caption{The execution time of the optimized models with different fault-tolerant DLA design strategies.}
\end{figure}

\noindent\textbf{Hardware resource overhead:} We evaluated the additional chip area overhead introduced by typical fault-tolerant DLA designs. This paper mainly focuses on fault tolerance in the computing array, and unless otherwise specified, the chip area overhead only considers the computing array part. For ease of comparison, the additional chip area overhead is normalized relative to the original unprotected computing array area, and the experimental results are shown in Figure 9. TMR-ALG essentially uses time-division multiplexing to provide redundant protection for sensitive layers and does not incur additional hardware cost. TMR-ARCH mainly utilizes space-division multiplexing at the architecture level to provide redundancy for sensitive layers, requiring a small number of controllers and selectors, resulting in a slight increase in area compared to Base. Although TMR-CRT only needs to protect the high-order bits, the specific location and quantization of the high-order bits are related, and the corresponding circuit logic area is large under the premise of unrestricted quantization, resulting in a high cost for circuit-level protection. TMR-CL takes into account the mutual influence of different layer parameters and significantly reduces the cost at the circuit level through quantization restrictions. On the other hand, by using the FlexHyCA architecture to separate and protect a small number of important neurons, although the area is larger than Base and TMR-ARCH, it is still acceptable relative to the benefits provided.
\begin{figure}[!t]
\centering
\includegraphics[width=2.5in]{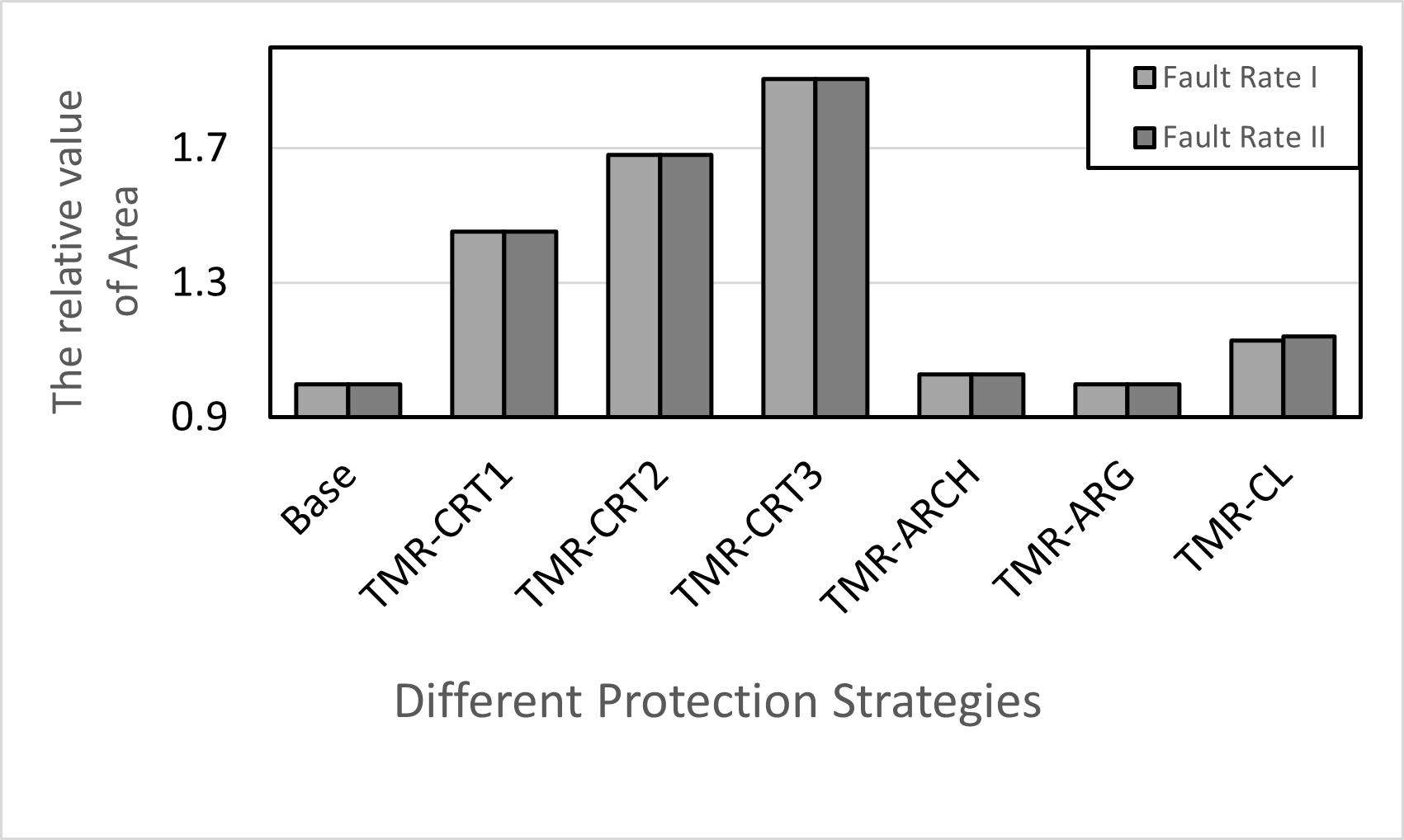}
\caption{To compare the relative area of chips corresponding to different fault-tolerant DLAs.}
\label{fig_9}
\end{figure}
Additionally, the TMR-CL is susceptible to the influence of uneven distribution of important neurons, which causes FlexHyCA to repeatedly load the dependencies of data and weights for DPPU, thereby introducing extra IO. Since other design schemes do not have extra IO growth, we will separately analyze the additional bandwidth consumption. The IO growth in TMR-CL mainly comes from two aspects: (1) when the proportion of important neurons in a computing unit exceeds the computing power of DPPU in a block calculation, we no longer reuse the data cached in the two-dimensional array, but directly load the required data from DRAM to avoid blocking the two-dimensional computing array. (2) We need to calculate the position of the two-dimensional array computing unit where the important neurons are located during the compilation phase so that FlexHyCA can selectively recompute. Finally, for the VGG16 and ResNet50 models, the additional IO introduced by TMR-CL is 8.2\% and 9.9\% relative to the weight data of the model itself, which will not have a significant impact on the overall DLA design's IO.
\subsection{Analysis of Algorithmic Layer Parameters}
In order to better understand the relationship between the core parameters of the algorithm, we analyze the relationship between the proportion of important neurons in the algorithm layer and the number of important bit positions in the neurons, using ResNet50 as an example. We also investigate the impact of model quantization constraints on model accuracy.

To study the relationship between the proportion of important neurons and the number of important bit positions, we set the proportion of important neurons to nine values: S\_TH={0.02, 0.05…0.3, ... , 0.35, 0.4}. We also considered six combinations of important bit positions for important neurons and regular neurons: {<2,1>, <3,1>, <4,1>, <3,2>, <4,2>, <4,3>}. The experimental results, shown in Figure 10, indicate that the proportion of important neurons has a significant impact on model accuracy when there are fewer bit protections. As the rate of important neurons increases, the accuracy also increases. However, in the presence of faults, the accuracy improvement shows a stage-like growth pattern. When the proportion of important neurons increases from 2\% to 5\%, there is a noticeable improvement in accuracy. However, when the proportion increases from 5\% to 20\%, the improvement is not significant. Only when the proportion increases to 25\% does the accuracy show a significant improvement. This suggests that the model can achieve high accuracy with only a small number of high bit positions and important neurons. However, further improvement in accuracy requires more bit positions and important neurons, which also means higher protection costs. When there are more bit protections, the effect of the proportion of important neurons becomes very weak. Essentially, this is because after the high bit positions of the deep learning model are fully protected, faults in the low bit parts have little impact on accuracy, whether they are important or regular neurons. However, the proportion of important neurons still plays a significant role in achieving ideal accuracy. Overall, when the user's requirements for reliability or accuracy are not very high, selecting a lower proportion of important neurons has a higher cost-effectiveness. For high accuracy requirements, selecting more bit protections has higher overall value, but we still tend to choose a lower proportion of important neurons.
\begin{figure}[htbp]
   \centering
   \captionsetup[subfloat]{labelfont=footnotesize,textfont=footnotesize}
   \subfloat[Fault rate I]{
    \includegraphics[width=3in]{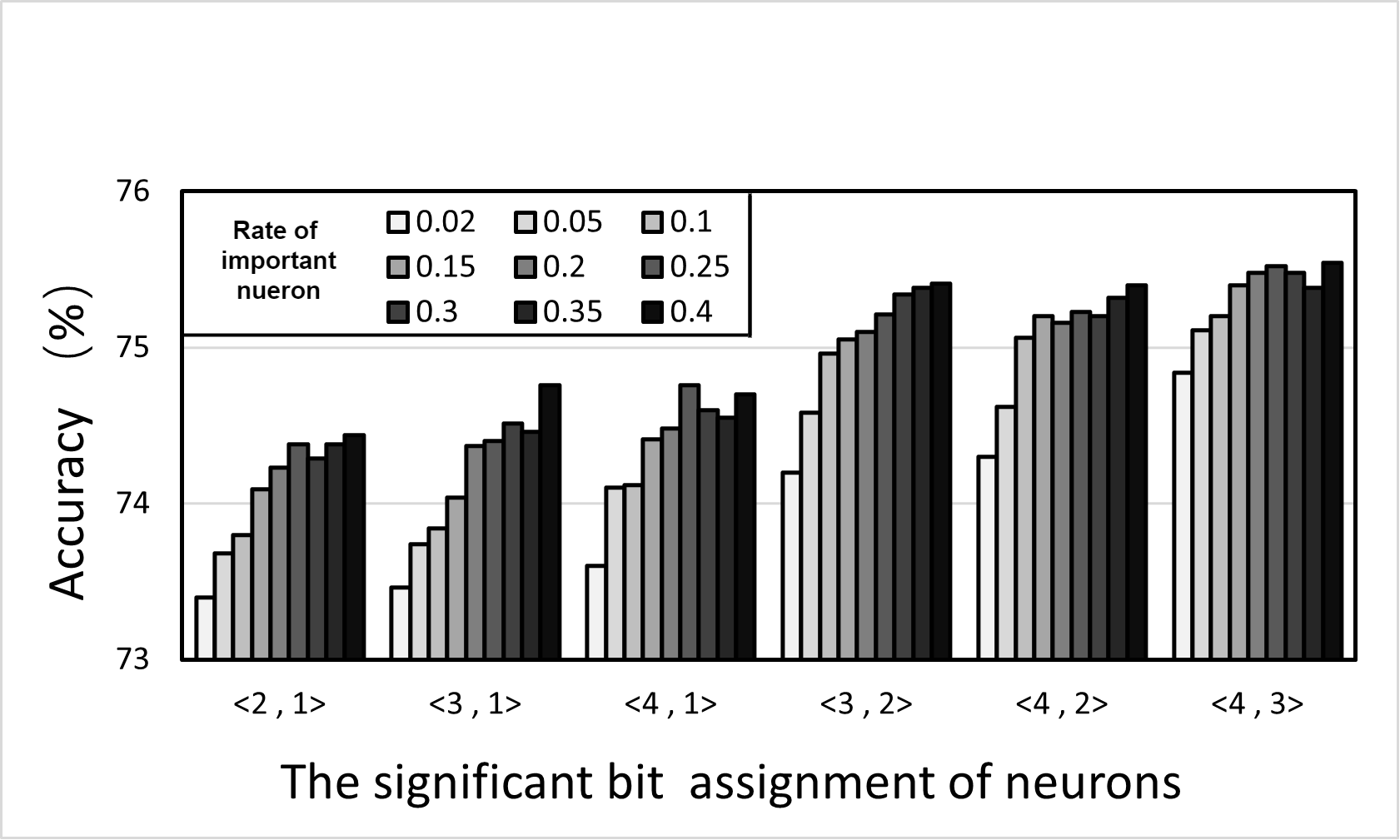}
  }
  \hfill
  \subfloat[Fault rate II]{
    \includegraphics[width=3in]{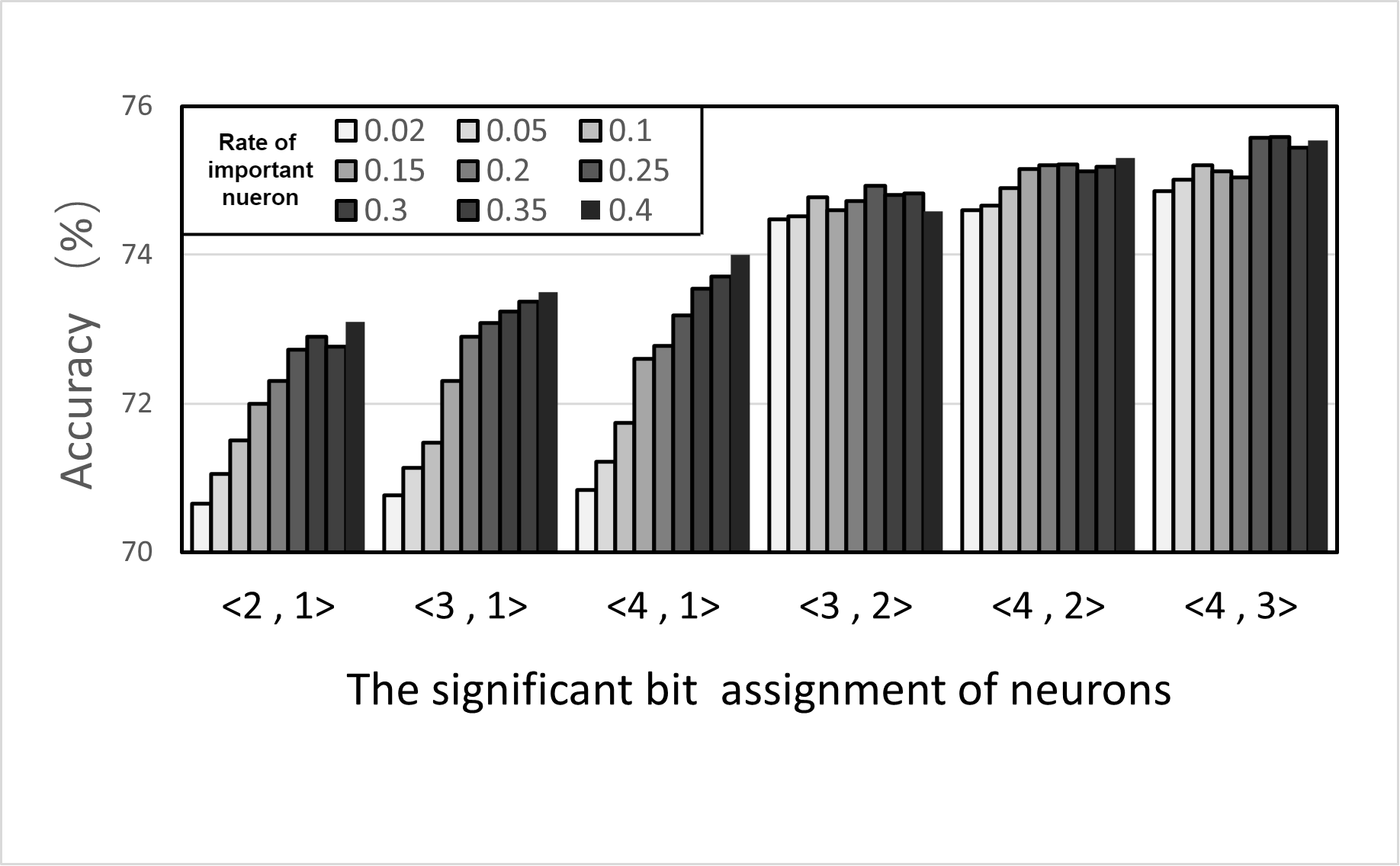}
  }
  \hfill
\caption{The relationship between the proportion of important neurons in ResNet50 and the number of important bit positions of neurons.}
\end{figure}

To simplify the analysis of the relationship between model quantization constraints and model accuracy, we applied a unified quantization constraint to the entire model. The accumulator data bit-width was set to 24 bits, and without constraints, the data range was any continuous position from bit 0 to bit 23. After adding the quantization constraint, we set the lowest bit of the truncated data as Q\_scale. Figure 11 shows the effect of Q\_scale on model accuracy, where the quantization selection space becomes smaller as Q\_scale increases, leading to a corresponding decrease in model accuracy. From the figure, it can be observed that the model accuracy drops very little when Q\_scale is less than 7. As discussed in Section 2.2, increasing Q\_scale can effectively reduce the logical circuit size corresponding to important bit positions. This means that within the allowed range of accuracy, there is still a significant amount of space to reduce the cost of selective bit protection.
\begin{figure}[!t]
\centering
\includegraphics[width=2.5in]{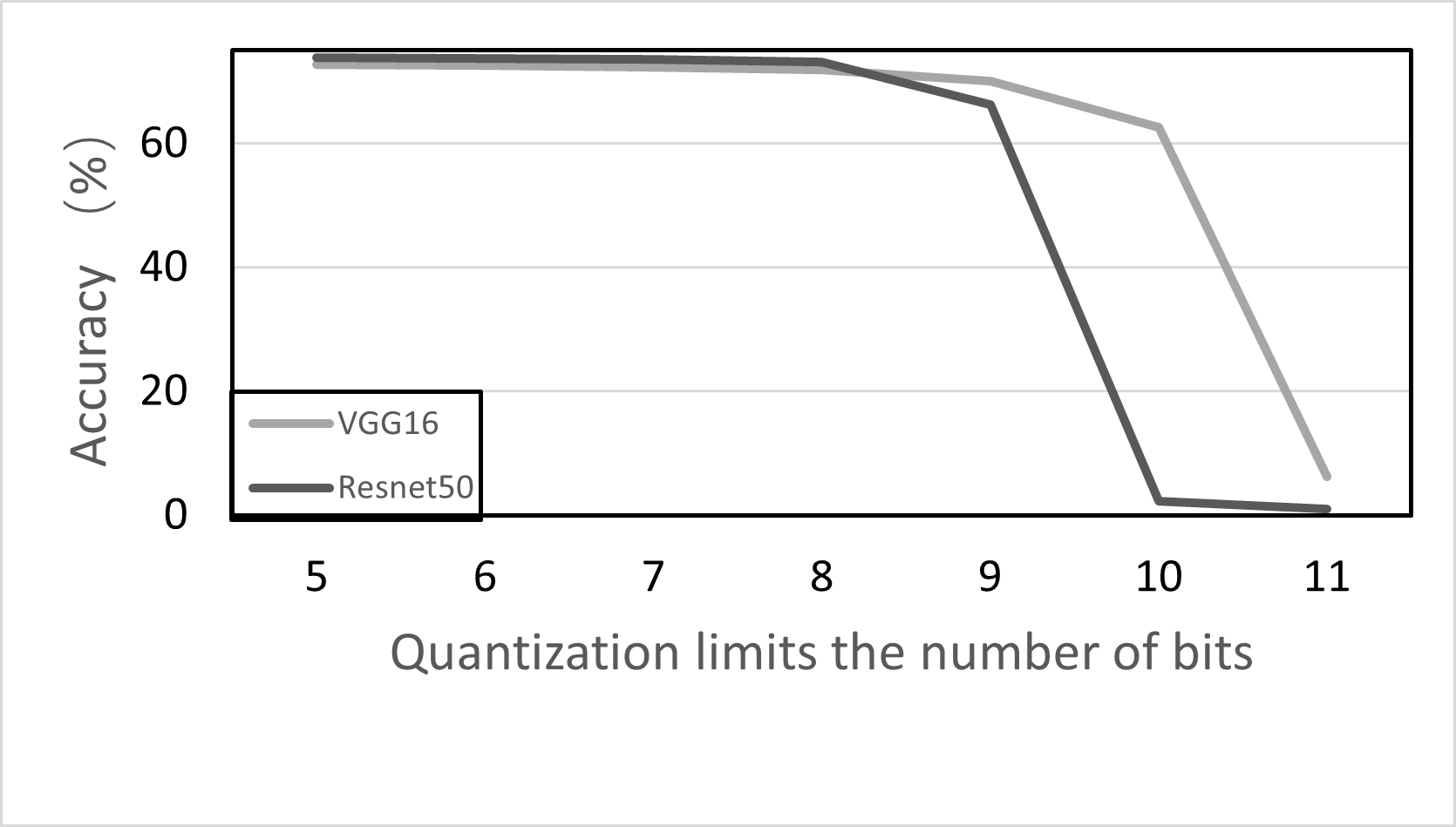}
\caption{The impact of the quantization constraint parameter Q-scale on the accuracy of the model.}
\label{fig_11}
\end{figure}
\subsection{Analysis of Architecture layer parameters}
In order to help understand the relationship between the main design parameters at the architecture level, we fixed the two-dimensional array of FlexHyCA to 32×32. Then, we analyzed the changes in chip area under different DPPU sizes and important bit settings, normalized to the chip area of the two-dimensional calculation array without protection. The experimental results are shown in Figure 12. Although the DPPU adopts more bit protection than the two-dimensional array, the number of calculation units in the DPPU is usually much smaller than that in the two-dimensional array. Therefore, the overall proportion of chip area introduced is relatively low. In contrast, the two-dimensional calculation array has a larger number of calculation units, and the introduction of chip area increases significantly with the increase of bit protection. Therefore, in the parameter optimization process, the bit protection setting of important neurons can be set larger, without causing a significant increase in chip area. In contrast, the bit protection setting of the two-dimensional calculation array should be kept as low as possible to control the overall redundant protection cost. In addition, if the DPPU is too large, it will inevitably introduce a larger chip area cost, but if it is too small, it will limit the proportion of important neurons. Essentially, the introduction of DPPU will only have good benefits if the selective protection provided by DPPU can cover the demand for bit protection of the two-dimensional array and the area cost of DPPU itself.
\begin{figure}[!t]
\centering
\includegraphics[width=2.5in]{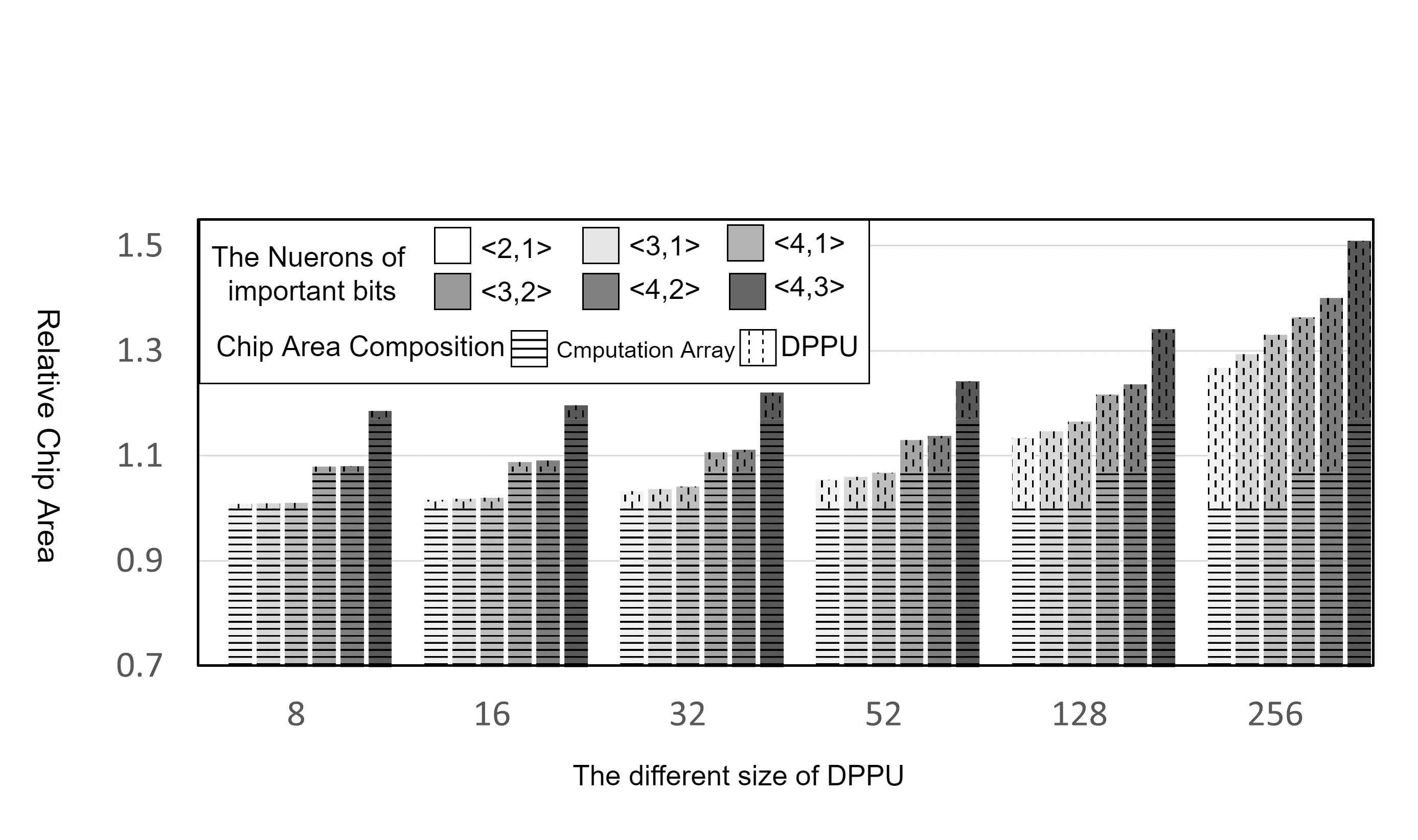}
\caption{The impact of DPPU size and bit protection settings on chip area}
\label{fig_12}
\end{figure}

We provide a direct data access path for DPPU, which avoids the blocking problem caused by the reuse of data in the two-dimensional array cache. However, this introduces additional memory access. Meanwhile, the design of FlexHyCA requires a significant amount of important neuron position information tables, which also leads to additional memory access. Therefore, we further evaluated the increase in I/O under different important neuron ratio settings and normalized it relative to the weight data. As shown in the figure, the additional DRAM data access linearly increases with the threshold design of important neurons. Thus, the additional data access records the positions of these important neurons. When the threshold design of neurons (S\_TH) is 0.1, the additional data loading exceeds 10\%, which does not meet our original design requirements. Therefore, the additional data loading also imposes certain limitations on the threshold design of neurons.
\begin{figure}[!t]
\centering
\includegraphics[width=2.5in]{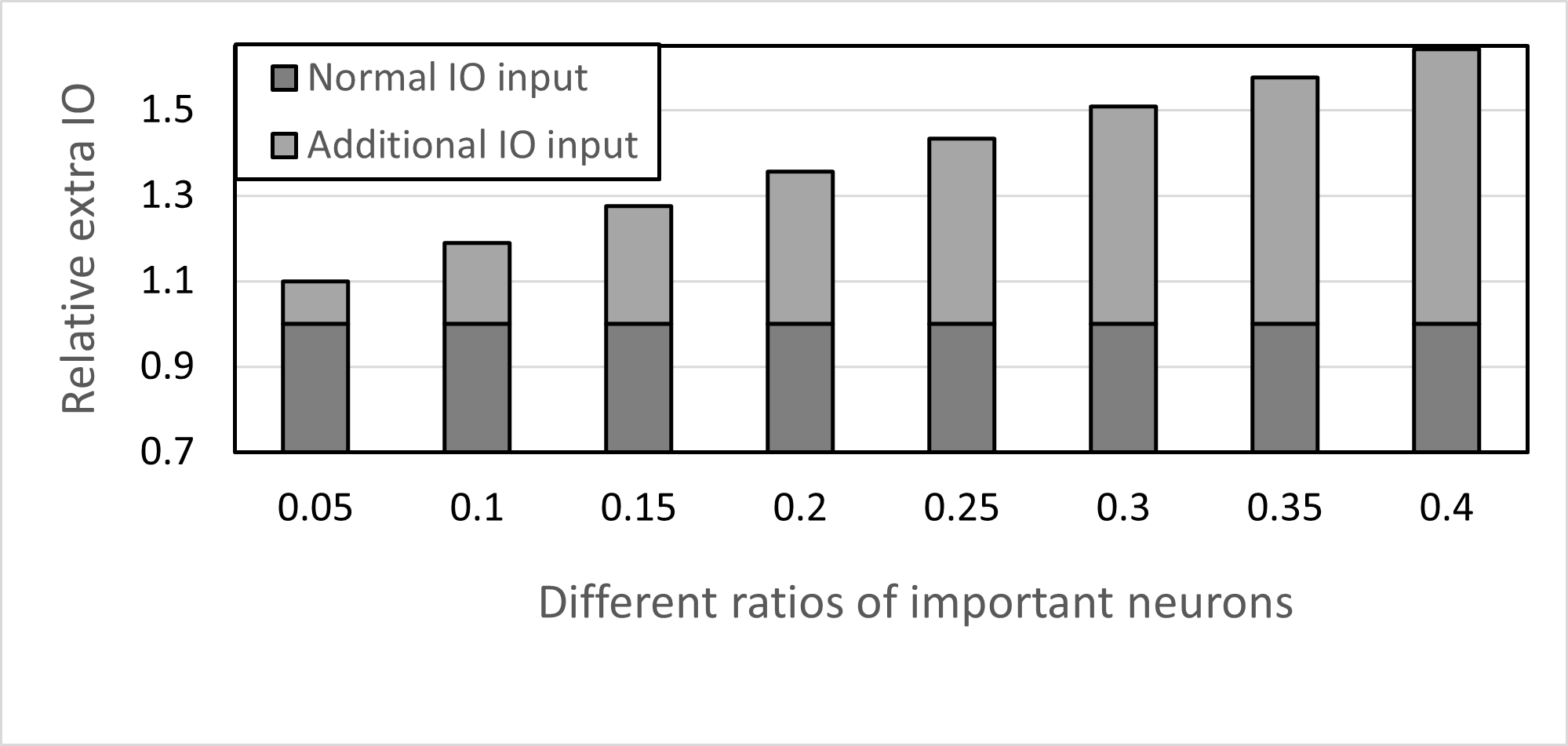}
\caption{The weight of important neurons has an impact on the input and output of FlexHyCA.}
\label{fig_13}
\end{figure}
\subsection{Analysis of Circuit-level Parameters}
To understand the parameters of circuit-level protection design, we analyzed the impact of different bit protections on the chip area of an 8-bit integer multiplier under unconstrained and constrained quantization, respectively. For the quantization constraint, we set Q\_scale to 4 and 7, based on the previous experiments where the impact of quantization constraint settings on the accuracy of ResNet and VGG was minimal. Additionally, for the constrained bit protection design, we further compared the direct redundancy and the reconfigurable redundancy implementation, where the direct redundancy method applies triple modular redundancy to the entire important logic area, while the reconfigurable redundancy uses MUX to selectively protect only the important bits according to the specific quantization. The experimental results are shown in Figure 14, which demonstrates that the area of fault-tolerant protection with constrained redundancy proposed in this paper is significantly reduced compared to the high-bit redundancy design without constraints. Compared to the simplest direct redundancy implementation, the redundancy area is reduced by an average of 71.4\%, mainly because only a few columns in the middle of the multiplier perform more calculations, resulting in higher redundancy costs. When we introduce quantization constraints, a small amount of constraint can bypass the redundancy of this part of the calculation, significantly reducing redundancy costs. Compared with direct redundancy, reconfigurable redundancy can reduce the number of redundancy calculation units, further reducing redundancy costs. Moreover, from the figure, it can be seen that the unconstrained redundancy strategy will also introduce significant area overhead when the number of bit protections is 1, because the logic scale of high-bit positions is still large under unconstrained quantization. In contrast, the cost of constrained redundancy becomes more stable with the increase of bit protection number, which is conducive to the scalability of the design.
\begin{figure}[!t]
\centering
\includegraphics[width=2.5in]{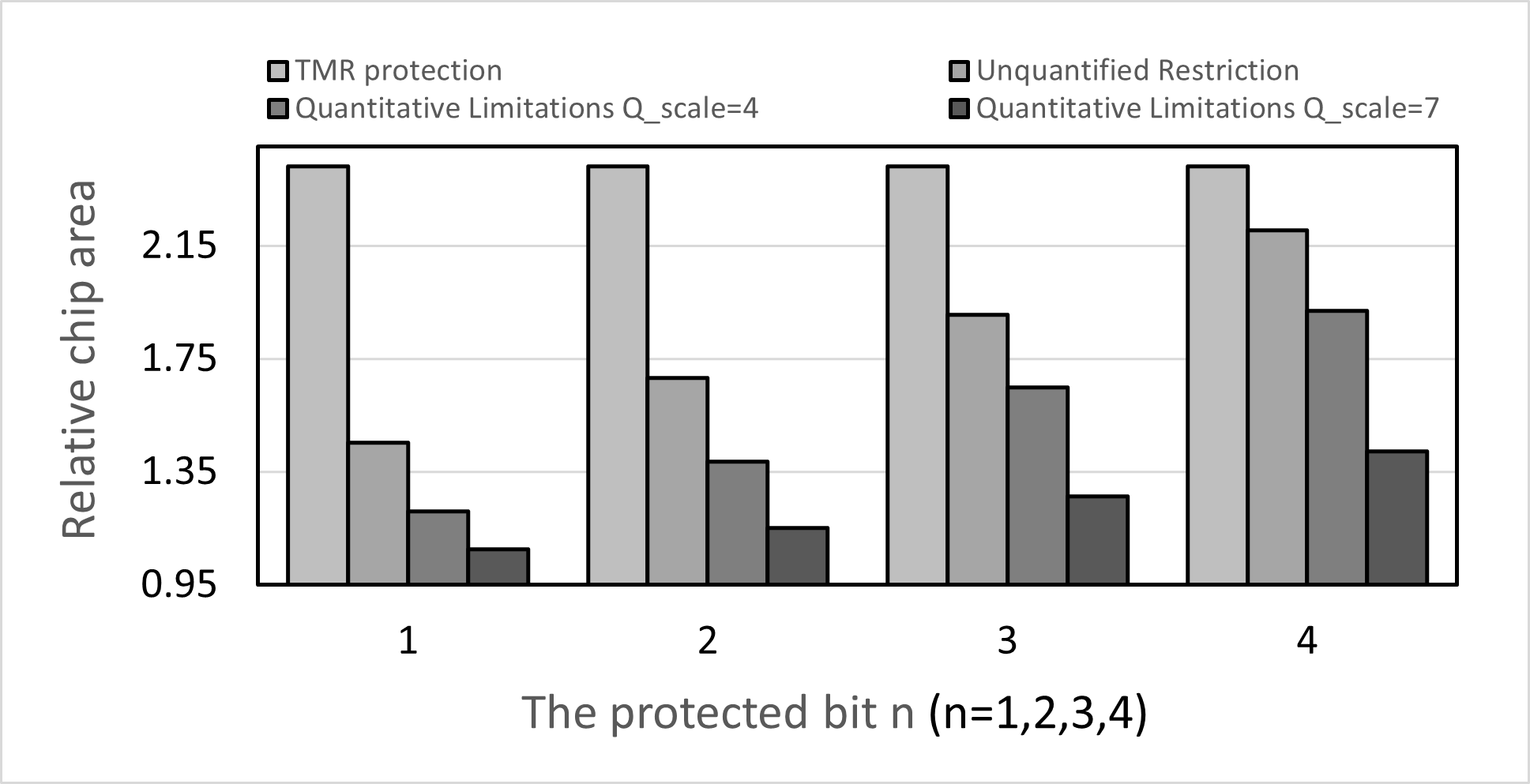}
\caption{The calculation array area corresponding to different bit protection designs.}
\label{fig_14}
\end{figure}
\subsection{Cross-Layer Design Parameter Search}
For the cross-layer design parameter search, we mainly use the Bayesian optimization algorithm to search for cross-layer design parameters in the design space. We aim to minimize the chip area  while satisfying the accuracy and performance constraints. The main design parameters and their ranges are shown in Table 1. Figure 14 shows the data points obtained using Bayesian search under different fault rates and their corresponding Pareto curves. Overall, it can be seen that the chip area 
 vary greatly for different design parameters. The parameter with the largest area cost corresponds to a chip area close to three times that of the optimized parameter. This also indicates the necessity of cross-layer parameter optimization. Meanwhile, from the figure, we can see that for both fault rates, when the model accuracy loss is around 3\% and 5\%, respectively, many data points in cross-layer optimization correspond to smaller chip areas. This is significantly advantageous compared to traditional triple modular redundancy. However, as the model accuracy requirement increases, the chip area cost corresponding to the data points obtained from cross-layer optimization increases significantly. Many design choices correspond to chip area costs that are more than twice the original chip area. This also indicates that after the accuracy requirement of fault-tolerant DLA increases, the corresponding fault-tolerant cost increases dramatically. Even if we only focus on the design choices on the Pareto curve, similar situations can be observed. When the reliability/accuracy requirement is low, the relative chip area cost can be controlled within 5\%. However, when the reliability/accuracy requirement increases, the optimized chip area cost can reach more than 40\%. Essentially, with the increase of accuracy requirement, the area redundancy cost increases gradually at first. This is because for the data points within this range, the high-one-bit protection is used in the two-dimensional computing array, and the threshold S\_TH\% for most important neurons is set relatively low (basically less than 10\%). Additionally, a configurable redundancy scheme is used. Under this scheme, the difference in area redundancy cost for protecting different bits of important neurons is not significant, and the area of DPPU is much smaller than that of the two-dimensional computing array. However, as the accuracy requirement further increases, the threshold or data bits that need to be protected for important neurons need to be further increased. The area of DPPU becomes significant compared to that of the two-dimensional computing array, and the required area redundancy cost increases significantly. Therefore, the reasonable requirements for deep learning reliability and accuracy have a significant impact on the final fault-tolerant design choice.

\begin{table}[!t]
\caption{Cross layer design parameter search space\label{tab:table1}}
\centering
\begin{tabular}{|c|c|c|}
\hline
\makecell[c]{Parameter\\variables}& Value & meaning\\
\hline
S\_TH &{5,10...35,40}  & \makecell[c]{Percentage of \\important neurons}\\
\hline
IN\_TH & {2,3,4}  & \makecell[c]{Important bits of \\important neurons} \\
\hline
NB\_TH & {1,2,3} & \makecell[c]{important bits of \\common neurons}\\
\hline
 Q\_scale & {1,2 ... 16} & Truncate constraints \\
\hline
S\_policy& \makecell[c]{layers\\sharding} & \makecell[c]{Selection strategies \\for important neurons}\\
\hline
Dot\_size & {8,16, ... ,256} & \makecell[c]{Points multiply \\array dimensions}\\
\hline
Data\_Reuse& {True,False} & \makecell[c]{Heterogeneous array \\data reuse}\\
\hline
PE\_policy &\makecell[c]{Direct\\Configurable}& \makecell[c]{Compute unit bit\\ protection policy} \\
\hline
\end{tabular}
\end{table}

In addition, we found that the collaborative design of quantization and circuit layers greatly reduces the fault-tolerant cost of the circuit layer. In scenarios where there is no distinction between important and normal neurons, the fault-tolerant cost is greatly reduced, and the space for cross-layer optimization is significantly narrowed. In the figure, we particularly marked the data points corresponding to high 1-bit protection, high 2-bit protection, and high 3-bit protection with squares. It can be seen that they are either on the Pareto curve or very close to it. Especially when many design parameters that meet the model accuracy requirements between high 1-bit protection and high 2-bit protection are shielded by the design of high 2-bit protection, a large blank space appears in the middle of the Pareto curve.

The corresponding optimized parameter results according to different experimental design goals are shown in Table 2. Overall, due to the strong correlation between many design parameters under different faults, such as the proportion of important neurons (IN\_TH) and the size of the dot product array (Dot\_size), the selection space for these two parameters is relatively small. This leads to some related parameter selection being affected, and as previously analyzed, the collaborative design of circuit layer and quantization greatly reduces the cost of circuit bit protection, significantly compressing the search space. On the other hand, the proportion of important neurons and the Dot\_size parameter are both discrete, which also limits the search space to a certain extent. These factors make the results of parameter search relatively close.
\begin{figure}[htbp]
   \centering
   \captionsetup[subfloat]{labelfont=footnotesize,textfont=footnotesize}
   \subfloat[Fault rate I]{
    \includegraphics[width=3in]{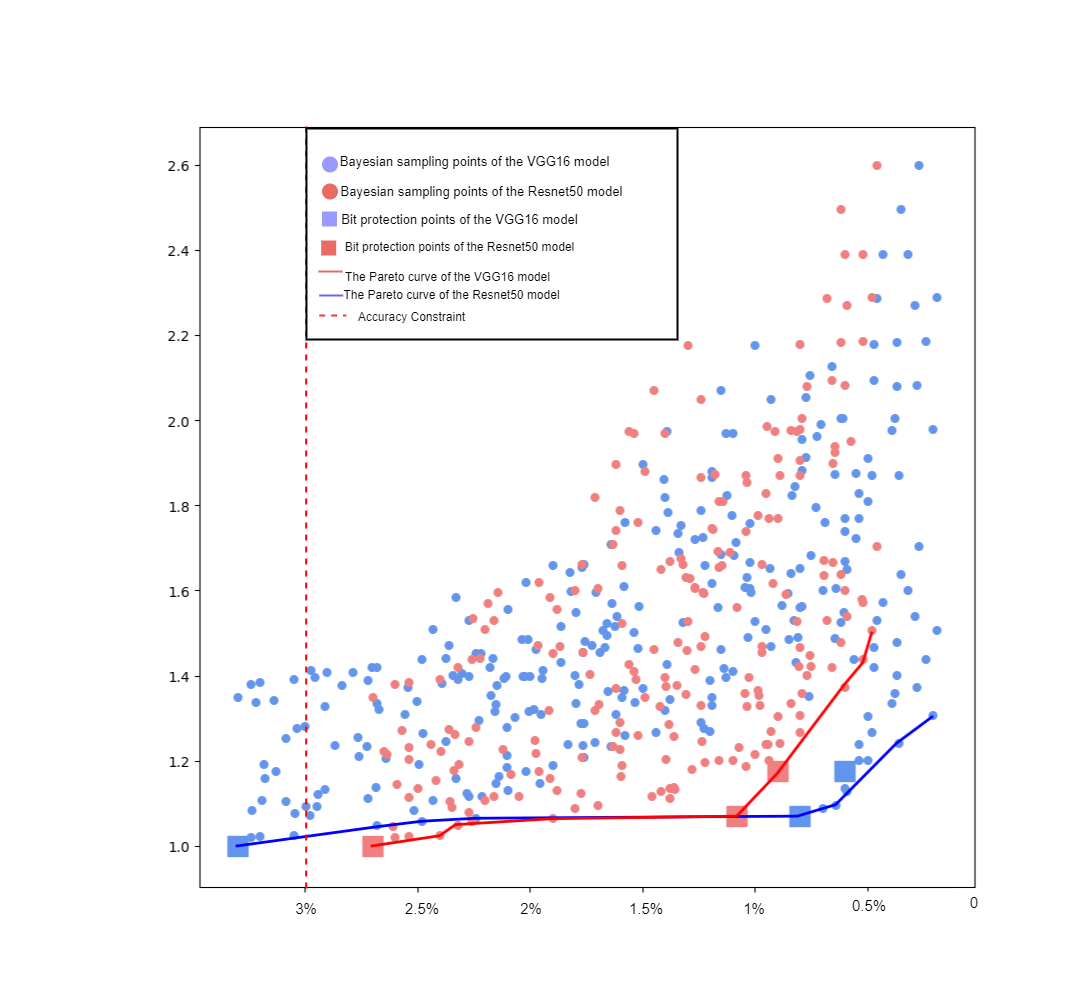}
  }
  \hfill
  \subfloat[Fault rate II]{
    \includegraphics[width=3in]{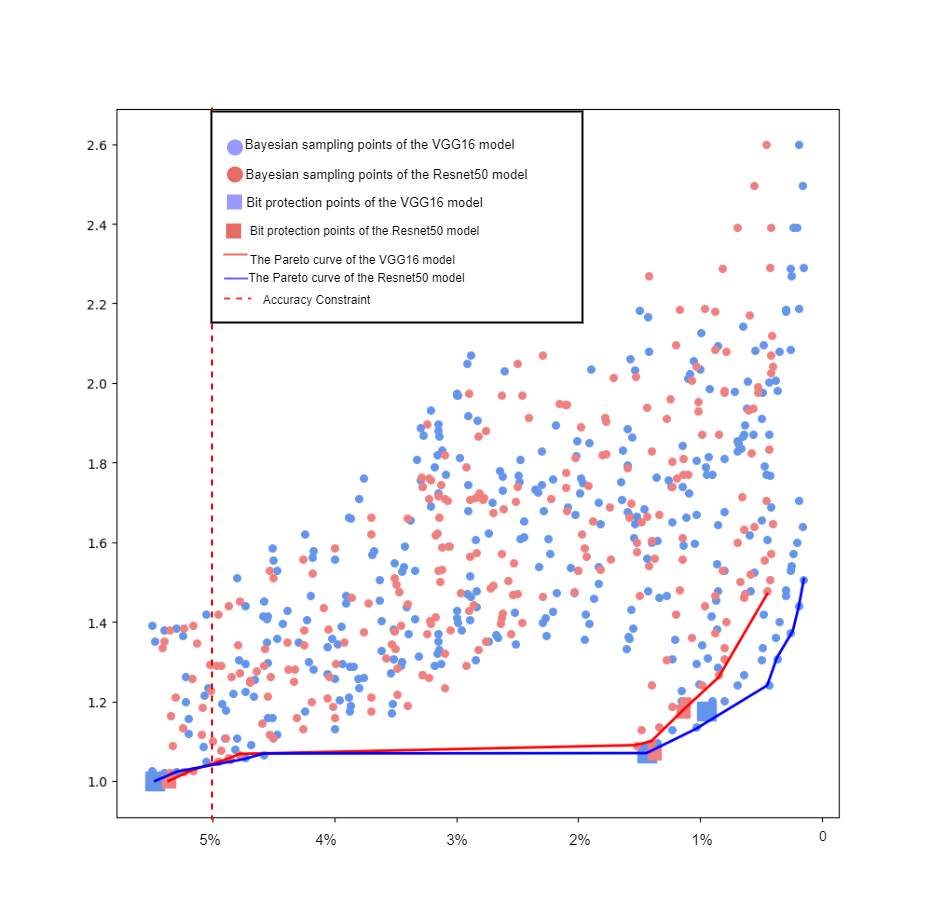}
  }
  \hfill
\caption{The data points sampled during the Bayesian optimization process.}
\end{figure}

\begin{table}[!t]
\caption{The best Cross layer design parameter of different model\label{tab:table2}}
\centering
\begin{tabular}{|c|c|c|}
\hline
Different Fault Rate & Fault I & Fault II\\
\hline
S\_TH & 5  & 5 \\
\hline
IN\_TH & 2  & 3 \\
\hline
NB\_TH & 1 & 1 \\
\hline
 Q\_scale & 7 & 8 \\
\hline
S\_policy& Uniform proportions& Uniform proportions\\
\hline
Dot\_size & 52 & 52 \\
\hline
Data\_Reuse& True & True \\
\hline
PE\_policy &Configurable &Configurable\\
\hline
\end{tabular}
\end{table}
\section{Conclusion}
Deep learning accelerators have become one of the mainstream computing engines for deep learning inference, and their reliability is a key factor in ensuring high-reliability deep learning. High-reliability deep learning accelerators have further increased the demand for reliability beyond traditional performance, energy efficiency, and area metrics, making the design space large and optimization difficult. To address the design of high-reliability deep learning accelerators, this paper proposes to explore the differences in the chip's sensitivity to soft faults for deep learning processing from two dimensions: neuron computation and neuron bit width, and based on these differences, propose corresponding selective protection methods from algorithm, architecture, and circuit design. At the same time, the relationship between cross-layer design parameters is systematically explored. Finally, with the help of Bayesian optimization, cross-layer parameter optimization is achieved, which can provide automated cross-layer fault-tolerant DLA design for different user requirements, minimize fault-tolerant protection costs while satisfying reliability/accuracy, performance, and other constraints. Experiments show that compared with traditional fault-tolerant designs that focus on a single layer, cross-layer fault-tolerant DLA better combines deep learning features with architecture and circuit design, greatly reducing the fault-tolerant cost of deep learning accelerators.

This paper systematically explores selective fault-tolerant protection methods for deep learning accelerators at the algorithm, architecture, and circuit levels. However, there are still several deficiencies that need to be addressed in the future in the following areas:

(1) The two-dimensional computational array in the accelerator architecture proposed in this paper currently cannot distinguish between regular neuron computation and important neuron computation. Essentially, important neurons are processed on both the two-dimensional array and the DPPU, which leads to computational waste and also limits the scale of the DPPU. In the future, sparse deep learning accelerator architecture can be used to remove important neuron computation from the two-dimensional computational array, further reducing fault-tolerant costs.

(2) Currently, we use a static method to separate important neuron computation from regular neuron computation, which lacks consideration of input data differences. In the future, we hope to further increase support for dynamic neuron importance, which will also help to further reduce fault-tolerant costs.

(3) There are still many other redundant protection methods at the algorithm level, such as fault-tolerant training, fault detection and correction based on checksum mechanisms, etc., especially in the case of low fault rates, which overlap with the fault-tolerant design strategies proposed in this paper. These methods can be incorporated into the cross-layer fault-tolerant design framework proposed in this paper, which is expected to obtain better fault-tolerant design solutions.

Author contributions statement: Author 1 completed the circuit design and conducted the experimental work, drafted the manuscript. Author 2 proposed the overall design scheme and revised the manuscript. Author 4 designed the algorithm for importance analysis and cross-layer parameter space exploration. Author 3, 5, and 6 provided guidance and feedback.

\end{document}